\documentclass[prd,aps,eqsecnum,floatfix,nofootinbib,preprint,tightenlines]{revtex4}
\usepackage{latexsym}
\usepackage{amsmath}
\usepackage{hyperref}

\def\bibi{\bibitem}

\def\a{\alpha}
\def\b{\beta}
\def\c{\chi}
\def\d{\delta}
\def\e{\epsilon}                
\def\g{\gamma}

\def\j{\psi}

\def\m{\mu}
\def\n{\nu}

\def\th{\theta}                  

\def\x{\xi}
\def\z{\zeta}
\def\D{\Delta}

\def\G{\Gamma}

\def\L{\Lambda}

\def\X{\Xi}

\def\cd{{\cal D}}

\def\cf{{\cal F}}

\def\cu{{\cal U}}
\def\cv{{\cal V}}

\def\cbo{{\,\raise-.15ex\Sc [\,}}                       
\def\ltap{\raisebox{-.4ex}{\rlap{$\sim$}} \raisebox{.4ex}{$<$}}   

\def\ddt#1{{\buildrel {\hbox{\LARGE .\kern-2pt.}} \over {#1}}}

\def\ie{\mbox{\it i.e.} }
\def\eg{\mbox{e.g.} }
\def\etc{\mbox{etc.} }

\def\leqx{\,\raisebox{-1.0ex}{$\stackrel{\textstyle <}{\sim}$}\,}

\def\tr{{\rm tr}\,}
\def\Tr{{\rm Tr}\,}

\def\Det{{\rm Det}}

\def\cO{{\cal O}}
\def\bI{\mbox{\boldmath $I$}}

\def\ie{{\it i.e.}}
\def\cf{{\it cf.}}
\def\eg{{\it e.g.}}
\def\etc{{\it etc}}

\def\qbar{{\overline{q}}}
\def\psibar{{\overline{\psi}}}
\def\chibar{{\overline{\chi}}}
\def\phibar{{\overline{\phi}}}
\def\Psibar{{\overline{\Psi}}}
\def\etabar{{\overline{\eta}}}

\def\etahat{{\hat{\eta}}}
\def\etahatbar{{\overline{\etahat}}}
\def\Hbar{{\overline{H}}}
\def\Psihat{{\hat{\Psi}}}
\def\Psihatbar{{\overline{\Psihat}}}
\def\qhat{{\hat{q}}}
\def\qt{{\tilde{q}}}
\def\etat{{\tilde{\eta}}}
\def\etatbar{{\overline{\etat}}}
\def\qhatbar{{\overline{\qhat}}}
\def\qtbar{{\overline{\qt}}}
\def\bltz{\mbox{\boldmath $B$}}
\def\Det{{\rm Det}}
\def\Tr{{\rm Tr}}
\def\tr{{\rm tr}}
\def\trts{{\tr_{ts}}}
\def\tD{{\tilde{D}}}
\def\tG{{\tilde{G}}}

\begin{document}
\hyphenation{fer-mio-nic per-tur-ba-tive pa-ra-me-tri-za-tion
pa-ra-me-tri-zed a-nom-al-ous}

\renewcommand{\thefootnote}{$*$}
\thispagestyle{empty}

\begin{center}
\vspace*{10mm}
{\large\bf Effective field theories for QCD with rooted staggered fermions}
\\[12mm]
Claude Bernard,$^a$ Maarten Golterman$^b$\ and\ Yigal Shamir$^c$
\\[8mm]
{\small\it
$^a$Department of Physics
\\Washington University,
Saint Louis, MO 63130, USA}
\\[5mm]
{\small\it
$^b$Department of Physics and Astronomy
\\San Francisco State University,
San Francisco, CA 94132, USA}
\\[5mm]
{\small\it $^c$Raymond and Beverly Sackler School of Physics and Astronomy\\
Tel-Aviv University, Ramat~Aviv,~69978~Israel}
\\[10mm]
{ABSTRACT}
\\[2mm]
\end{center}

\begin{quotation}
Even highly improved variants of lattice QCD with staggered
fermions show significant violations of taste symmetry at
currently accessible lattice spacings.  In addition, the
``rooting trick'' is used in order to simulate with the correct
number of light sea quarks, and this makes the lattice theory
nonlocal, even though there is good reason to believe that
the continuum limit is in the correct universality class.  In
order to understand scaling violations, it is thus necessary
to extend the construction of the Symanzik effective theory to
include rooted staggered fermions.  We show how this can be
done, starting from a generalization of the renormalization-group
approach to rooted staggered fermions recently developed by
one of us.  We then explain
how the chiral effective theory follows from the Symanzik
action, and show that it leads to ``rooted'' staggered chiral
perturbation theory as the correct chiral theory for QCD with rooted
staggered fermions.   We thus establish a direct link between the
renormalization-group based arguments for the correctness of the
continuum limit and the success of rooted staggered chiral perturbation
theory in fitting numerical results obtained with the rooting trick.
In order to develop our argument,
we need to assume the existence of a standard partially-quenched
chiral effective theory for any local partially-quenched theory.
Other technical, but standard, assumptions are also required.

\end{quotation}

\renewcommand{\thefootnote}{\arabic{footnote}} \setcounter{footnote}{0}

\newpage
\section{\label{intro} Introduction}
On a hypercubic lattice in four dimensions, the continuum limit of
lattice QCD with staggered fermions \cite{ks} contains four ``tastes''
of mass-degenerate quarks per staggered fermion field
\cite{kasm,stw,gs84,saclay}.\footnote{
  We assume the usual choice of only a single-site bare mass term.
}
Hence, if we introduce a separate staggered fermion field
for each physical light-quark flavor (up, down, and strange), the continuum
limit consists of QCD containing four up, four down, and four
strange quarks.

A simple solution to this problem is to adjust for the
excessive multiplicity by taking
the fourth root of the fermion determinant for each staggered
fermion field \cite{parisi}.  Heuristically, if the staggered determinant
factorizes into four identical determinants in the continuum limit,
one for each taste, taking the fourth root corrects for the
taste multiplicity.
The desired theory, QCD with one up, down and strange
quark each is then obtained in the continuum limit.
Since the staggered determinant
is positive for any real, nonzero bare quark mass $m$, and the continuum determinant
is (formally) positive for positive quark mass, the positive
fourth root should be chosen.\footnote{
  For the case of an odd number of quarks with negative quark mass, see
Refs.~\cite{dh06,bgss06}.
}
The continuum quark mass is proportional to $|m|$,
which undergoes only a multiplicative renormalization,
because staggered fermions have one exact chiral symmetry.

This procedure, the ``fourth-root trick,'' raises a number of
questions \cite{reviews,sharpelat06,bgslat06}.  The fourth root
of a determinant cannot in general be written as a Grassmann integral
with a local action.
Therefore, the first question is whether the theory defined by the
fourth-root trick is local and unitary.

In Ref.~\cite{bgs06} we showed that, as might be expected, the fourth-root
staggered theory is not local at nonzero lattice spacing $a$.
Continuing correlation functions defined in the Euclidean
theory to Minkowski space will lead to violations of
unitarity at $a\ne 0$, on a distance scale set by the lightest particles
in the theory, the Goldstone bosons.
For examples of this, see Ref.~\cite{SP}, as well as Sec.~6 of Ref.~\cite{bernard06},
which we will revisit later in this paper.

The origin of these
diseases can be traced back to the taste symmetry-breaking part
of the staggered Dirac operator.  This taste-breaking part
corresponds to a dimension-five irrelevant operator.
Thus, in the local, unrooted staggered theory, all taste symmetry-breaking
effects are expected to vanish in the continuum limit,
where exact $U(4)$ taste symmetry will be restored
for each of the four up, four down, and four strange quarks
present in that theory.

The leading power-law scaling of irrelevant operators
is characteristic of any local and renormalizable theory,
such as in particular the unrooted staggered theory.
This brings us to the second question: Does the same scaling persist
in the fourth-root theory?
Two related considerations make it natural
to address this question via a Renormalization-Group (RG)
approach.
To begin with, the RG framework
allows us to define what we mean by the continuum limit.
This is done by performing $n+1$ blocking steps\footnote{
  See Sec.~\ref{SET} for an explanation of the convention $a_c/a_f=2^{n+1}$
  \cite{shamir06}.
}
on the original lattice theory,
with its fine spacing $a_f=a$, each time increasing the lattice spacing
by a factor of two, to arrive at an RG-blocked theory
formulated on a lattice with a coarse spacing $a_c=2^{n+1}a_f$.
Keeping $a_c$ fixed and small in physical units, $a_c\ll\L_{QCD}^{-1}$,
while sending $n\to\infty$ (and thus $a_f\to 0$),
one obtains a coarse-lattice theory describing the continuum physics.
An RG framework is also natural because the restoration of taste
symmetry is only expected to occur on distance scales
much larger than the original lattice cutoff $a_f$.
RG blocking removes the short-distance fluctuations while modifying
the action of the remaining degrees of freedom by local terms only.
When we increase the number of blocking steps $n$,
the blocked theory becomes more taste symmetric, and we eventually recover
exact taste symmetry in the continuum limit $n\to\infty$.

Using this RG framework, it was argued in Ref.~\cite{shamir06} that
the continuum limit of QCD with rooted staggered fermions
is a local theory that belongs to the correct universality class.
There are strong arguments that the
fourth-root theory, while nonlocal, is nevertheless renormalizable
\cite{bg94,sharpelat06,giedt06}, and this is the fundamental reason behind
the validity of its continuum limit.
The detailed reasoning is based on a number of technical assumptions,
all of which are very similar to the assumptions needed
to establish the nature of the continuum limit for
the {\it unrooted} staggered theory.
Further analytic and numerical
work aimed at confirming the technical assumptions of Ref.~\cite{shamir06}
would add direct and strong evidence for the validity of the fourth-root trick.
For full details, we refer to Ref.~\cite{shamir06};
for shorter, more intuitive accounts,
we refer to Refs.~\cite{bgslat06,sharpelat06}.
We stress that one key element---the anticipated scaling of
the taste-breaking effects---has been corroborated by extensive
numerical studies \cite{cbmilc06,milc,FM,evs}.

Assuming that the rooted staggered theory has the correct continuum limit,
this leaves us with a third question.  While the anticipated scaling
of taste-breaking effects is observed,  these effects are clearly
not negligible at present \cite{cbmilc06,milc,FM,SP,evs}.
It is therefore imperative to take lattice artifacts into account
in the effective continuum field theories
(EFTs) such as the Symanzik effective theory
(SET) or chiral perturbation theory (ChPT).
The latter provides a central tool for analyzing the numerical data
and performing the chiral and continuum extrapolations in the light-quark
sector.  In the case of rooted staggered fermions,
we thus need to construct EFTs that take
the discretization effects
into account,
including those that correspond to the nonlocal
behavior of the theory at $a\ne 0$.
The construction of such EFTs is the subject of this paper.

For the pseudo-scalar Goldstone-boson physics,
a candidate EFT already exists;
it is provided by staggered ChPT \cite{ls99} with the replica rule (rSChPT), or
``rooted staggered ChPT'' \cite{ab03}.
(Extensions to higher order \cite{Sharpe:2004is}, and to
heavy-light meson \cite{ab05} and baryon \cite{bailey} rSChPT
were recently given.)
An argument for the validity of rSChPT
was presented in Ref.~\cite{bernard06}, and reviewed in Refs.~\cite{bgslat06,sharpelat06}.
The key feature of Ref.~\cite{bernard06} is that the argument
takes place completely within the context of chiral effective theories,
and the replica rule is justified only in that context.  Here we will need to
introduce a somewhat different version of the replica rule, which
will be justified in addition at the level of the fundamental lattice
theory, but which will ultimately give the same results in the chiral theory.
A detailed comparison of the two approaches will be made in Sec.~\ref{comparison}.

The overall goal in the current paper is
to extend the standard procedure for the construction of
ChPT for a local lattice theory to QCD with rooted staggered fermions.
The standard procedure consists of two steps.
The SET \cite{symanzik} is constructed first.  This can be done order by order
in perturbation theory, but it is generally assumed that the SET is
valid nonperturbatively as well.
We will assume throughout that this includes partially quenched theories
\cite{bg94}.
In particular, we will assume that locality suffices,
and that unitarity (which may be lost in partially quenched theories) is not necessary.
Once the correct form of the SET has been established,
its symmetries can be used to construct ChPT.  Since the SET organizes the
low-energy effective theory as a systematic expansion in the lattice spacing, one
automatically obtains the chiral theory as an expansion in the lattice
spacing as well.

Establishing that EFTs can be constructed following the
usual rules for QCD with rooted staggered fermions thus constitutes
a fundamental step in understanding the effects of rooting at nonvanishing lattice spacing.
The main thrust of this paper is the construction of the SET
for the rooted theory; obtaining the corresponding ChPT
is then straightforward, and we show that it is indeed given by rSChPT.
We emphasize that our construction applies to
all commonly used versions of staggered fermions:
standard (unimproved) staggered \cite{ks}, Asqtad \cite{ASQTAD}, HYP \cite{HYP}, Fat7bar \cite{FAT7BAR}, HISQ
\cite{HISQ}, {\it etc}.
The only requirement is that the action have the usual staggered symmetries.
The size of the discretization effects is of course different with
different versions of staggered fermions, but their form (and appearance at each order in $a_f$) is the same.

It is also important to note that the
effective theories we ultimately construct are those for the relevant rooted staggered theory on
the original (fine) lattice.  The RG framework is used only as a tool in the derivation of
these effective theories.  Nevertheless, it is an indispensable tool:
the conclusions of Ref.~\cite{shamir06} have to be valid in order for our construction of the EFTs to make any sense.
We will assume this to be the case.

The difficulty in constructing EFTs for the rooted theory is the following.
Consider for simplicity a staggered theory with a common power, denoted $n_r$,
of the fermion determinant for each staggered flavor in the theory.
As long as $n_r$ is a positive integer the lattice theory is local,
and the construction of EFTs proceeds as usual.
In order to arrive at the fourth-root theory,\footnote{
  The discussion generalizes easily to the isospin limit
  $m_u=m_d\equiv m_\ell$, where one takes the square root of
  a single staggered flavor with (bare) mass $m_\ell$.
}
however, we must set $n_r=1/4$.
Our task is to ensure that a \textit{replica continuation}
may be performed:  a well-defined procedure
must be devised to reach the value $n_r=1/4$ at the level of an EFT,
and the procedure must be
consistent with the $n_r$-dependence of the underlying lattice theory.

In a diagrammatic EFT calculation, the dependence on the number
of (sea) quarks arises in two ways.  First, there is explicit dependence arising through
loop diagrams. In addition, the coupling constants of the EFT
(the Symanzik coefficients in the case of the SET, and the low-energy
constants in the case of ChPT) depend in an unknown way
on the underlying lattice theory,
including in particular on the number of replicas $n_r$.
It is the latter dependence that makes our task nontrivial.
In principle, one may envisage two basic obstructions
to the replica continuation of the coupling constants in the EFT.
Mathematically, a unique analytic continuation off the positive
integers (which in the case at hand is where the theory is local)
does not exist.  Also, it could be that the replica continuation
we have in mind will encounter a singularity precisely at the desired point $n_r=1/4$.

The dependence of the underlying lattice theory
on the number of replicas $n_r$
is both perturbative and nonperturbative; this means that
proving that no obstacle to the replica continuation is present
would be tantamount to solving the theory nonperturbatively.
The key observation that makes our task nevertheless tractable is that,
after a large number $n$ of RG blocking steps,
the taste-symmetry breaking effects are very small:
the unrooted staggered theory with integer $n_r$ is very close to a $U(4)$
taste-invariant theory.  The rooted theory, with
$n_r=1/4$, is then also very close to a \textit{local} lattice theory,
for which the standard construction of EFTs is valid.
Indeed, the ``re-weighted'' taste-invariant theories
introduced in Ref.~\cite{shamir06} are local whenever $n_r$ is a multiple
of $1/4$.  The proximity of these local theories
makes it possible to construct the SET and, later, ChPT,
for the rooted theory.

We will reach the SET for the rooted theory starting from
the SET for the corresponding re-weighted, taste-invariant theory.
The flavors of the taste-invariant theory will always be kept
in one-to-one correspondence
with those of the continuum-limit theory.
In the taste-invariant theory the dependence of the
Symanzik coefficients on the physical quarks is nonperturbative, and unknown,
as usual.  This does not pose any difficulty, because
the number of physical flavors is never varied.

During the intermediate steps of the derivation
the parameter $n_r$ will take on a related, but different technical meaning.
The precise definitions will be given and explained in Sec.~\ref{SET} below.
As already mentioned above,
we first approximate the staggered theory by a local, taste-invariant
theory that belongs to the correct universality class.
The (rooted) staggered theory will then be reached from the taste-invariant
theory by ``turning on'' smoothly the taste-breaking effects.
The dependence on $n_r$ of the lattice theory
will come only from the taste-breaking effects, which are
nonlocal (for noninteger $n_r$) but small.
The difference between corresponding taste-invariant and staggered theories
is of order the fine lattice spacing $a_f$ of the original (unblocked)
lattice.  This will allow us to show that
all the lattice correlation functions are polynomials in $n_r$
to any fixed order in the expansion in $a_f$.  The degree of
the $n_r$-polynomial is less than the order of the $a_f$-expansion.
The $n_r$-dependence of the Symanzik coefficients can then be determined
unambiguously. It follows that the replica continuation
is nowhere singular in the complex $n_r$ plane, to the given order in $a_f$.
Finally, after performing the replica continuation,
the parameter $n_r$ resumes its original role as the
power of the staggered determinant in the lattice theory.
The further transition to ChPT is
essentially a repeat of the same reasoning.  As will become clear later on,
we do have to assume that a chiral effective theory can be constructed for
any local, but partially-quenched, theory.  This was already emphasized in
Refs.~\cite{bernard06,sharpelat06}.

The outline of this paper is as follows.  In Sec.~\ref{symm} we
consider the symmetries of staggered fermions in some detail.
We derive the form in which shift symmetry \cite{gs84} is
realized in the SET, and thus in any other EFT derived from the SET.
A quick overview of the most important observations of that section
is given at its beginning, and any reader not interested in the details
can skip the remainder of the section.

In Sec.~\ref{SET} we come to the main part of this paper,
the construction of the SET for QCD with rooted staggered fermions.
We generalize the staggered theory to a class of partially-quenched theories
in which it is possible to implement the program outlined above.
In Sec.~\ref{examples} we discuss the SET to quadratic order in the
lattice spacing in more detail, in order to illustrate the general construction.
In Sec.~\ref{SChPT} we make the transition to the chiral effective theory,
and demonstrate that it is indeed given by rSChPT.  As an example, we work
out in rSChPT the leading-order contribution
to the connected scalar two-point function, following the
calculation in Ref.~\cite{bernard06}. We then compare the present
derivation of rSChPT to that of Ref.~\cite{bernard06}, using the respective
discussions of the scalar two-point function to make the comparison concrete.
The final section contains our conclusions.
A brief account of this work was presented at Lattice~2007 \cite{bgslat07}.

\section{\label{symm} Symmetries of the Symanzik effective action for staggered fermions}
Here we discuss the symmetries of unrooted staggered fermions that are most relevant for
this paper, and the way they appear at the level of the SET.
We begin with an overview of the main results of this section.  In the following subsections
we will then give a more detailed discussion.

\begin{itemize}
\item[1.]
The staggered fermion action is invariant under shift symmetry, which, in the continuum
limit, enlarges to the product of $SU(4)$ taste symmetry and translation
symmetry.  At the level of the SET,
the taste part of shift symmetry takes the form of the
32-element group $\G_4$ generated by a set of four-dimensional
Dirac gamma matrices $\x_\m$, with
\begin{equation}
\label{Diracalg}
\{\x_\m,\x_\n\}=2\d_{\m\n}\ ,\ \ \ \ \m,\n\in\{1,2,3,4\}\ .
\end{equation}
This result was derived to order $a^2$ in Ref.~\cite{ls99}. Here we give a
general argument that makes it clear that the result is true to all orders
in $a$.  On the continuum quark fields $q$ used in the SET,
the generating elements of $\G_4$ can be chosen to act according to
\begin{equation}
\label{shiftSET}
q\to\x_\m q \ ,\ \ \ \ \qbar\to\qbar\x_\m\ .
\end{equation}
Here the field $q_{\b b}$ has a Dirac
spin index $\b$ and an $SU(4)$ taste index $b$, with the matrices $\x_\m$ acting on the latter.

\item[2.]
On the lattice, a taste-basis field $\j$
carrying the same indices as the continuum quark field $q$
is related to the one-component field $\chi$ by a
unitary transformation \cite{saclay,gliozzi}
\begin{equation}
\label{basistransf}
\psi=Q\chi\ ,\ \ \ \ \ \psibar=\chibar Q^\dagger\ .
\end{equation}
The field $\j$ lives on a coarse lattice whose spacing is twice that
of the original staggered action.
The ultra-local, unitary matrix $Q$ maps the one-component variables $\c$
on the sixteen sites of each even hypercube to the sixteen components
of $\j$ on the single corresponding coarse-lattice site.
The transformation $Q$ is required to be gauge covariant,
and its choice is not unique.  As a result hypercubic rotational symmetry
is somewhat complicated in the taste basis.\footnote{
  For a detailed discussion of rotational symmetry in this framework,
  see Ref.~\cite{shamir06}.
}
Of course, since the one-component and taste bases are related by
a unitarity transformation, the
physical consequences of all staggered symmetries are preserved.

A somewhat different taste-basis operator, that we will refer to as the
``RG taste-basis'' Dirac operator
\cite{shamir05,shamir06}, is defined by a Gaussian smearing of
the unitary transformation (\ref{basistransf}).
The resulting inverse Dirac operator satisfies
\begin{equation}
\label{RGDirac}
D_{taste}^{-1}=\frac{1}{\a}+QD_{stag}^{-1}Q^\dagger\ ,
\end{equation}
where $D_{stag}$ is the Dirac operator in the one-component formulation,
and $\a$ is a parameter of order $1/a$.  Even though the theories described
by $D_{stag}$ and $D_{taste}$ are no longer related by a simple,
unitary basis transformation, they are physically completely
equivalent, because the propagators differ only by a contact term.
The advantage of the Gaussian-smeared transformation
is that discarding the taste-breaking part of $D_{taste}$
does not introduce any fermion doublers \cite{shamir05,bgs06}.
Because the staggered theory and the taste-invariant theory have
a similar fermion content, one can interpolate smoothly between them.
This will prove useful for the derivation of the SET.

In the one-component formulation,
shift symmetry is a unitary transformation on the fields
$\chi$ and $\chibar$ (\cf\ Eq.~(\ref{shift}) in the next subsection).
Since $Q$ is unitary, the same is also true for
the fields $\psi$ and $\psibar$,
and from this it follows that the theory in the RG taste basis is also
invariant under shift symmetry.

\item[3.]
Because of staggered symmetries, discretization errors for theories
with staggered fermions start at order $a^2$ \cite{Sharpe:1993ng}.
This is not obvious if one considers
staggered fermions in the taste basis of Refs.~\cite{saclay,gliozzi},
or in the modified form used in the
RG analysis of Refs.~\cite{shamir05,shamir06}, where taste-breaking
terms occur in the action starting at order $a$.
In this case, shift symmetry connects
the leading, taste-invariant term in the lattice action
with the order $a$ taste-breaking term, \ie, their relative strength is fixed.
There exists a local field redefinition that brings the taste-basis
lattice action into a form where the taste violations are explicitly
of order $a^2$, and shift symmetry is realized as in Eq.~(\ref{shiftSET}),
again up to order $a^2$ terms \cite{luo}.
More generally, the momentum-space basis used in the derivation of
Eq.~(\ref{shiftSET}) can be related to the taste basis by a non-local
field redefinition.  Because the construction of the SET proceeds order by
order in $a$, the field redefinition in effect becomes local. Therefore,
the SETs constructed in the taste basis and in the staggered
(or momentum-space) basis are always related by a local field redefinition.

\item[4.]
Staggered fermions have an exact chiral symmetry when $m=0$,
often referred to as $U(1)_\e$ symmetry, taking the form \cite{kasm}
\begin{equation}
\label{U1eps}
\chi(x)\to e^{i\th\e(x)}\chi(x)\ ,\ \ \ \
\chibar(x)\to e^{i\th\e(x)}\chibar(x)\ ,\ \ \ \ \
\e(x)=(-1)^{x_1+x_2+x_3+x_4}\ .
\end{equation}
For $m=0$, this implies that
\begin{equation}
\label{GWrel}
\{D_{taste},\g_5\otimes\x_5\}=\frac{2}{\a}D_{taste}(\g_5\otimes\x_5)D_{taste}\ ,
\end{equation}
where $\g_5$ acts on the spin index,
and $\x_5=\x_1\x_2\x_3\x_4$ acts on the taste index
\cite{shamir05,bgs06}.
In other words, $D_{taste}$ is a Ginsparg--Wilson operator \cite{gw}
with respect to $U(1)_\e$ symmetry.

\end{itemize}

Before we proceed, we return to the relation of our analysis
and that of Ref.~\cite{ls99}.  The SET at order $a^2$ was determined
in Ref.~\cite{ls99} by enumerating the
allowed dimension-6 lattice operators consistent with the lattice symmetries,
including shift symmetry.  It was then shown that shift symmetry
is represented on the corresponding continuum operators as a
$\G_4$ symmetry.  A more direct method of determining the SET,
which we follow here, is to enumerate continuum operators.
This leads to the result of point~1, that shift symmetry always
implies a taste $\Gamma_4$ symmetry of the SET.

In the subsections following below, we will discuss some of these observations in more
detail.  These subsections are not needed for the construction of the SET for rooted staggered
fermions, which can be found in Sec.~\ref{SET}.

\subsection{\label{diagram} Diagrammatic argument}
Our first argument for claim 1 above is essentially perturbative,
and assumes that we are working in the momentum-space representation
of the one-component basis.  This result may be considered a corollary
of Ref.~\cite{gs84}.  To keep it self-contained, a summary of relevant
facts from Ref.~\cite{gs84} has been included in the discussion below.

We will consider diagrams with $n$ external fermion
and $r$ external gauge-field lines,
corresponding to an operator which appears at a certain
order in the SET.
On the lattice, because of the phase factors which appear in the staggered action,
momentum is conserved modulo $\pi$ (in this section we work in lattice units),
and any such diagram will have an overall
delta function for momentum conservation of the form
\begin{equation}
\label{delta}
\delta(p_1+\dots+p_n+k_1+\dots+k_r+\Pi)\ ,
\end{equation}
where $\Pi$ is a vector with components $0$ or $\pi$.
The delta function is the
periodic delta function with period $2\pi$.
The (lattice) quark and anti-quark momenta are $p_i$,
$i=1,\dots,n$ and the gluon momenta $k_j$ , $j=1,\dots,r$.

Because we are interested in an operator in the SET,
we may take all physical external momenta small.  Fermion doubling
then implies that on every quark line we need to split the momenta as
\begin{equation}
\label{splitmom}
p_i=q_i+\pi_{A_i}\ ,
\end{equation}
in which $q_i$ lives in the reduced Brillouin zone ($-\pi/2<q_{i\mu}\le\pi/2$), and $\pi_{A_i}=\pi A_i$,
with
\begin{equation}
\label{A}
A_i\in\{(0,0,0,0), (1,0,0,0), \dots, (1,1,1,1)\}\ .
\end{equation}
We now take all physical momenta
$q_i$ and $k_j$ small --- so small that their sum has no components as large as
$\pm\pi$.  The delta function in Eq.~(\ref{delta}) thus factorizes into
\begin{equation}
\label{factor}
\delta(q_1+\dots+q_n+k_1+\dots+k_r)\;\delta(\pi_{A_1}+\dots+\pi_{A_n}+\Pi)\ .
\end{equation}

Now consider what happens to this diagram under a shift
\begin{eqnarray}
\label{shift}
\chi(x)&\to&\zeta_\mu(x)\chi(x+\hat\mu)\ ,\\
\chibar(x)&\to&\chibar(x+\hat\mu)\zeta_\mu(x)\ ,\nonumber\\
U_\nu(x)&\to &U_\nu(x+\hat\mu)\ ,\nonumber\\
\zeta_\mu(x)&=&(-1)^{x_{\mu+1}+\dots+x_4}=e^{i\pi_{\z_\m}\cdot x}\ ,\nonumber
\end{eqnarray}
where the last equality defines $\pi_{\z_\m}$.
In momentum space (with $\chi(x)=\int_p e^{ip\cdot x}\chi(p)$), this
takes the form
\begin{equation}
\label{shiftmom}
\chi(p_i)=\chi(q_i+\pi_{A_i})\to e^{i(q_i+\pi_{A_i})_\mu}\chi(q_i+\pi_{A_i}+\pi_{\zeta_\mu})\ .
\end{equation}

Applying a shift in the $\m$ direction to all external legs of our diagram, and noting that the $j$th external gluon line
is multiplied by a factor $e^{i(k_j)_\m}$ under a shift, we obtain the total
factor
\begin{equation}
\label{smallmom}
e^{i(q_1+\dots+q_n+k_1+\dots+k_r)_\mu}\ ,
\end{equation}
which, by virtue of the first delta function in Eq.~(\ref{factor}), is equal to one.  Therefore,
we may omit these (small-momentum) phase factors in the shift~(\ref{shiftmom}).
We conclude that the diagram is invariant under the modified symmetry
\begin{equation}
\label{shiftmommod}
\chi(q_i+\pi_{A_i})\to e^{i(\pi_{A_i})_\mu}\chi(q_i+\pi_{A_i}+\pi_{\zeta_\mu})\ ,
\ \ \ \ \ i=1,\dots,n\ ,
\end{equation}
which does not act on the gluon fields.
The transformation (\ref{shiftmommod}) generates a representation
of the group $\G_4$ acting on the quark fields.
Indeed, applying the transformation first in the $\mu$ direction, and then in the
$\nu$ direction, one obtains (dropping the index $i$)
\begin{equation}
\label{munu}
\chi(q+\pi_A)\to e^{i(\pi_A+\pi_{\zeta_\mu})_\nu}\;e^{i(\pi_A)_\mu}\chi(q+\pi_A+\pi_{\zeta_\mu}+
\pi_{\zeta_\nu})\ .
\end{equation}
For $\mu=\nu$, we have $(\pi_{\zeta_\mu})_\mu=0$ (\cf\ Eq.~(\ref{shift})), and Eq.~(\ref{munu})
thus reduces to the identity.  For $\mu\ne\nu$,
\begin{eqnarray}
\label{signs}
\zeta_\mu(x+\nu)=\zeta_\mu(x)\ \Rightarrow\  e^{i(\pi_{\zeta_\mu})_\nu}&=&+1\ ,\ \ \ \ \ \mu>\nu\ ,\\
\zeta_\mu(x+\nu)=-\zeta_\mu(x)\ \Rightarrow\  e^{i(\pi_{\zeta_\mu})_\nu}&=&-1\ ,\ \ \ \ \ \mu<\nu\ ,\nonumber
\end{eqnarray}
implying that shifts anti-commute, just like the generators of $\G_4$.
We may make contact with Eq.~(\ref{shiftSET})  by introducing
\begin{equation}
\label{redef}
\phi_A(q)\equiv\chi(q+\pi_A)\ .
\end{equation}
The transformation Eq.~(\ref{shiftmommod}) can now be written as
\begin{equation}
\label{Xi}
\phi_A(q)\to \sum_B (\Xi_\mu)_{AB}\phi_B(q)\ ,
\end{equation}
for some $16\times 16$ matrices $\Xi_\mu$ satisfying the Dirac algebra
\begin{equation}
\label{Dirac}
\{\Xi_\mu,\Xi_\nu\}=2\delta_{\mu\nu}\ .
\end{equation}
Finally, we can perform a basis transformation such that $\X_\m=1\otimes\x_\m$,
and transform back to position space to obtain Eq.~(\ref{shiftSET}).

Our argument shows that the diagram is invariant under the symmetry
(\ref{shiftSET}) if it is invariant under shift symmetry (\ref{shiftmom}).
The group $\G_4$ may thus be used to restrict the form of the SET
in accordance with the shift symmetry of the underlying lattice theory.
This is a considerable simplification, because the group $\G_4$
does not mix operators
of different dimensions, \ie, of different orders in the Symanzik expansion.

The same reasoning goes through in a theory in which the staggered
fermion fields carry a flavor index $\ell=1,\dots,n_f$: one simply labels the fields $\chi_\ell$ and $\chibar_\ell$
in Eq.~(\ref{shift}) with the extra index $\ell$.   Since the gauge fields also transform under
shift symmetry, the same shift symmetry acts on all staggered fields
simultaneously.  It thus follows that the discrete symmetry $\G_4$ acts in the same
way on all staggered fields $\chi_\ell$, and does not enlarge to the group $(\G_4)^{n_f}$
\cite{ab03}.

As an aside, we note that the invariance of the diagram under shift
symmetry has implications for the second delta function in Eq.~(\ref{factor}).
Naively, it would seem
to follow that $\Pi$ just has to be equal to the sum over all $\pi_{A_i}$,
but in general this is not sufficient.
The reason is that the vertex can contain explicit periodic functions
of the external momenta, which leads to additional sign factors under a shift.  This is
best illustrated with an example.  Consider a lattice vertex of the form
\begin{equation}
\label{vertexexample}
\sum_{A,B}\delta(q_1+q_2+k)\delta(\pi_A+\pi_B+\Pi)\cos{(q_1+k+\pi_A)_\nu}
\chibar(q_2+\pi_B)\chi(q_1+\pi_A)A_\nu(k)\ ,
\end{equation}
in which we split $p_1=q_1+\pi_A$, $p_2=q_2+\pi_B$, and take $q_{1,2}$ and $k$ to be small.
Performing a shift on the $\chi$ and $\chibar$ fields
results in (dropping a factor $\delta(q_1+q_2+k)$)
\begin{eqnarray}
\label{exshifted}
&&\hspace{-1cm}\sum_{A,B}\delta(\pi_A+\pi_B+\Pi)\cos{(q_1+k+\pi_A)_\nu}
\;e^{i(\pi_A+\pi_B)_\mu}\;\\
&&\hspace{5cm}\times\
\chibar(q_2+\pi_B+\pi_{\zeta_\mu})\chi(q_1+\pi_A+\pi_{\zeta_\mu})A_\nu(k)\nonumber\\
&&\hspace{-1cm}=
\sum_{A,B}\delta(\pi_A+\pi_B+\Pi)\cos{(q_1+k+\pi_A)_\nu}\;e^{i(\pi_{\zeta_\mu})_\nu}
\;e^{i\Pi_\mu}\;
\chibar(q_2+\pi_B)\chi(q_1+\pi_A)A_\nu(k)\ ,\nonumber
\end{eqnarray}
where we used that $(\pi_{\zeta_\mu})_\mu=0$ and that $2\pi_{\zeta_\mu}=0\ {\rm mod}\ 2\pi$.
The vertex is thus invariant if $\Pi_\mu+(\pi_{\zeta_\mu})_\nu=0\ {\rm mod}\ 2\pi$.
An example of such a $\Pi$ is $\pi_{\eta_\nu}$, which is defined by the phase
factors which appear in the staggered action:
\begin{equation}
\label{eta}
\eta_\nu(x)\equiv e^{i\pi_{\eta_\nu}x}\equiv (-1)^{x_1+\dots+x_{\nu-1}}\ .
\end{equation}

\subsection{\label{group} Group-theoretical argument}
There is a very simple group-theoretical way to derive the
same result.  Let $S_\mu$ be the shift in the $\mu$ direction.  All
elements of the shift-symmetry group can be generated from the
basic four shifts, and it is thus sufficient to consider only the $S_\mu$.
In any irreducible representation of the group, $S_\mu$ looks like
\begin{equation}
\label{shiftsymm}
S_\mu\to e^{iq_\mu}\Xi_\mu\ ,
\end{equation}
with $-\pi/2< q_\mu\le\pi/2$ the physical momentum in
lattice units, and the matrices $\Xi_\mu$ generate a representation of $\G_4$
\cite{golterman86}.
All irreducible representations are either ``bosonic,'' if each $\Xi_\mu$
is mapped onto $\pm 1$ (sixteen choices), or ``fermionic,'' if
the $\Xi_\mu$ are chosen to satisfy the Dirac algebra~(\ref{Dirac}).
Any field appearing in an EFT for the staggered theory (such as the SET or ChPT)
transforms in some
representation of $S_\mu$ under a shift (\ie, with some choice of
$q_\mu$ and $\Xi_\mu$).

Now we use that any continuum EFT is also invariant under continuum translations,
which, on a continuum field $\Phi$ with momentum $q$, act as
\begin{equation}
\label{transl}
\Phi(q)\to e^{iq\cdot r}\Phi(q)\ ,
\end{equation}
for a  translation over a displacement $r$.  We may thus choose
$r$ such that $q\cdot r = -q_\mu$, follow $S_\mu$
by this translation, and again obtain a symmetry of the EFT.
This symmetry is precisely the one generated by the $\Xi_\mu$,
\ie, a representation of $\G_4$.

\subsection{\label{taste} Taste basis}
The arguments in the previous subsections made use of the momentum
basis of the one-component formalism.  The
Feynman rules for the staggered theory in the one-component basis \cite{gs84}
were (assumed to have been) used in the derivation of the SET.
Also, the group-theoretical argument works naturally on
the momentum basis, since that is where irreducible representations
of the staggered symmetry group live \cite{golterman86}.
Alternatively, one could have started from
the taste basis.  The SET derived from the taste basis will not look
the same as that derived from the one-component formalism;
but the two SETs should be physically equivalent.  Since the one-component
and taste bases are related by a (nonlocal) unitary transformation in
momentum space \cite{saclay}, one expects that the SETs derived from them, too,
will be related by a field redefinition. Moreover,
to any finite order in $a$, the SET-level field redefinition should be local,
because the same is true for the unitary transformation between the two bases,
when expanded to the corresponding finite order in $a$.

We illustrate this in the free massless theory, working to order $p^2$ in the
Symanzik expansion.  On the taste basis,
shift symmetry takes on the form \cite{saclay,luo}
\begin{equation}
\label{tasteshift}
\psi(y)\to\frac{1}{2}\left((\xi_\mu+\g_5\g_\mu\xi_5)\psi(y)
+(\xi_\mu-\g_5\g_\mu\xi_5)\psi(y+\hat\mu)\right)\ .
\end{equation}
The field $\psi$, introduced in Eq.~(\ref{basistransf}),
is in this case given explicitly by
\begin{equation}
\label{stag-to-taste}
\psi_{\beta b}(y) = \frac{1}{2^{3/2}}\sum_A (\g_A)_{\beta b}\chi(2y+A)\ ,
\end{equation}
where
$\g_A=\g_1^{A_1}\g_2^{A_2}\g_3^{A_3}\g_4^{A_4}$, and
$A$ runs over the set (\ref{A}).  The normalization factor in Eq.~(\ref{stag-to-taste})
differs from that in Ref.~\cite{saclay} because we take $\psi$ to be in lattice
units of the coarser lattice; whereas Ref.~\cite{saclay} works in physical units.
In momentum space, Eq.~(\ref{tasteshift}) looks like
\begin{eqnarray}
\label{tsmom}
\psi(p)&\to&e^{ip_\mu/2}\left(\xi_\mu\cos{(p_\mu/2)}-i\g_5\g_\mu\xi_5\sin{(p_\mu/2)}\right)\psi(p)\\
&=&e^{ip_\m/2}\left(\x_\m-\frac{i}{2}\g_5\g_\m\x_5p_\m+\cO(p^2)\right)\psi(p)\ .\nonumber
\end{eqnarray}
The factor $e^{ip_\mu/2}$ corresponds to the factor $e^{iq_\mu}$
in Eq.~(\ref{shiftmom}), because the lattice spacings differ by a factor two.
Dropping the factor $e^{ip_\m/2}$ on the same grounds as in Sec.~\ref{diagram},
it is easily verified that the transformation (\ref{tsmom})
becomes a generating element of $\G_4$,
and that it is a symmetry of the order-$a$ SET in the taste
representation,
\begin{equation}
\label{SETtaste}
S_{free}=\sum_\mu\int_{-\pi}^{\pi}
\frac{d^4p}{(2\pi)^4}\;
 \psibar(p)\left(i\g_\mu p_\mu+\frac{1}{2}\,\g_5\xi_5\xi_\mu p_\mu^2+\cO(p^3)\right)\psi(p)\ .
\end{equation}
This may also be written as
\begin{equation}
\label{SETtrace}
S_{free}=\sum_\mu\int_{-\pi}^{\pi}
\frac{d^4p}{(2\pi)^4}\;
 \left(\tr\left[\psibar(p)i\g_\mu p_\mu\psi(p)\right]+\frac{1}{2}\,\tr\left[\psibar(p)\g_5 p_\mu^2\psi(p)(\g_5\g_\mu)^\dagger\right]
+\cO(p^3)\right)\ ,
\end{equation}
where we consider the field $\psi_{\b b}$ as a $4\times 4$ matrix.

In momentum space, the transformation relating the one-component and taste
representations is \cite{saclay,daniel,luo}
\begin{eqnarray}
\label{fieldredef}
\psi(p)&=&\frac{1}{2^{11/2}}\sum_{A,B}(-1)^{A\cdot B}\g_A\,\phi_B(q)\,e^{iq\cdot A}\ ,\\
\psibar(p)&=&\frac{1}{2^{11/2}}\sum_{A,B}(-1)^{A\cdot B}\g_A^\dagger\,\phibar_B(q)\,e^{-iq\cdot A}
\ ,\nonumber
\end{eqnarray}
where again $A$ and $B$ take values in the set (\ref{A}), and where $q=p/2$.
The field $\phi(q)$ was defined in Eq.~(\ref{redef}).
The transformation (\ref{fieldredef}) is indeed nonlocal,
but its expansion to any finite order in $a$ is local.  For instance,
upon expanding $e^{\pm iq\cdot A}=1\pm iq\cdot A+\cO(q^2)$, and starting
from Eq.~(\ref{SETtrace}),
this field redefinition brings the action~(\ref{SETtaste}) into the form
\begin{equation}
\label{SEToc}
S_{free}=\sum_\m\sum_{AB}\int_{-\pi/2}^{\pi/2}
\frac{d^4q}{(2\pi)^4}\;
\phibar_A(q) \left(i(\G_\m)_{AB}\,q_\m
+\cO(q^3)\right)\phi_B(q)\ ,
\end{equation}
where the $\G_\m$ matrices
form a 16-dimensional representation of the Dirac algebra
and commute with the taste matrices $\X_\n$ defined in Eq.~(\ref{Xi}).  Note that
Eq.~(\ref{SEToc}) is expressed in
units of the fine lattice spacing.

Let us also briefly consider the RG taste representation defined by
Eq.~(\ref{RGDirac}) in the massless free theory. To order $p^2$ the action
is given by \cite{bgs06}
\begin{equation}
\label{RGtaste}
\sum_\mu\int_{-\pi}^{\pi}
\frac{d^4p}{(2\pi)^4}\;
 \psibar(p)\left(i\g_\mu p_\mu+\frac{1}{\a}\;p_\m^2+\frac{1}{2}\,\g_5\xi_5\xi_\mu p_\mu^2+\cO(p^3)\right)\psi(p)\ .
\end{equation}
This action is invariant under $U(1)_\e$ symmetry in ``Ginsparg--Wilson--L\"uscher'' (GWL)
\cite{gw,luescher98}
form.  In the free theory, this symmetry looks like (again to order $a$)
\begin{eqnarray}
\label{U1epstaste}
\delta\psi(p)&=&\g_5\xi_5\left(1-\frac{2}{\a}\sum_\mu i\g_\mu p_\mu+\cO(p^2)\right)\psi(p)\ ,\\
\delta\psibar(p)&=&\psibar(p)\;\g_5\xi_5\ .\nonumber
\end{eqnarray}
In this case, we may first carry out a field redefinition
\begin{eqnarray}
\label{GWfieldredef}
\psi(p)&\to&\left(1+\frac{1}{\a}\sum_\mu i\g_\mu p_\mu\right)\psi(p)\ ,\\
\psibar(p)&\to&\psibar(p)\ ,\nonumber
\end{eqnarray}
followed by (\ref{fieldredef}), to bring the action into a form without
terms of order $a$.
Note that Eq.~(\ref{GWfieldredef}) is nothing but the free-field, order-$a$
form of the field redefinition
\begin{equation}
\label{GWfrint}
\psi\to(1-D/\a)^{-1}\psi
\end{equation}
with here $D=D_{taste}$,
which transforms the GWL
form of $U(1)_\e$ symmetry into a standard $\g_5\xi_5$ symmetry \cite{gs00}.

As a final note, we observe that to this order in $a$, field redefinitions can be
carried out such that the resulting action is invariant under the full $U(4)$ taste
symmetry.   This turns out to be true to all orders in $a$ in the free theory \cite{gs84}, but not in the
interacting theory.

\section{\label{SET} Derivation of the Symanzik effective action}
We begin by considering a theory with $n_r$ replicas of one staggered fermion
with bare mass $m$, in the RG taste basis.  For now, $n_r$ will be a positive
integer.
We perform $n+1$ RG blocking steps, labeled $k=0,1,\ldots,n$,
following the blocking procedure of Ref.~\cite{shamir06}.  The special $k=0$
step is used to carry out the transition from the one-component to the taste basis,
\cf~Eq.~(\ref{RGDirac}).  In this step the number of fermion degrees
of freedom is not thinned out; in each subsequent step they are thinned out by a factor
$2^4=16$.
The partition function for this theory can be written as
\begin{equation}
\label{RGstag}
Z(n_r)=\int\cd\cu\prod_{k=0}^n\cd\cv^{(k)}\;
\bltz_n\left(n_r;\cu,\{\cv^{(k)}\}\right)\;
\Det^{n_r}\left(D_{taste,n}\right)\ .
\end{equation}
The notation here is as follows:  The gauge field on the original
lattice, with spacing $a_f$, is denoted by $\cu$.
The spacing of the $k$-th blocked lattice is $a_k=2^{k+1}a_f$,
and the gauge field on that lattice is $\cv^{(k)}$.
The spacing of the final, coarse lattice is  $a_c=2^{n+1}a_f$.
The Boltzmann weight for the collection of gauge fields, original
and blocked, is $\bltz_n\left(n_r;\cu,\{\cv^{(k)}\}\right)$.
It is composed of three parts: the original gauge action,
the gauge-field blocking kernels,\footnote{
  We do not integrate over any of the gauge fields;
  this can be postponed to the end.
  The explicit expression for $\bltz_n\left(n_r;\cu,\{\cv^{(k)}\}\right)$
  is given in Ref.~\cite{shamir06}.
}
and a short-distance contribution to the effective gauge-field action,
$n_r\,\d S_{eff}$, coming from
integrating out the fermions on all lattices except the last one, where
\begin{equation}
\label{RGshort}
e^{-\d S_{eff}}=\prod_{k=0}^n\Det\left(G_k^{-1}\right)\ .
\end{equation}
The operators $D_{taste,k}$ and $G_k^{-1}$ are recursively given by
\begin{eqnarray}
\label{RGrecursion}
D_{taste,k}^{-1}&=&\a_k^{-1}+Q^{(k)}D_{taste,k-1}^{-1}Q^{(k)\dagger}\ ,
\ \ \ \ \ k=1,\dots,n\ ,\\
G_k^{-1}&=&D_{taste,k-1}+\a_k Q^{(k)\dagger}Q^{(k)}\ ,
\ \ \ \ \ k=1,\dots,n\ .\nonumber
\end{eqnarray}
The blocking parameter $\a_k$ is of order $1/a_k$.
The blocking kernel at the $k$-th step, $Q^{(k)}=Q^{(k)}(\cv^{(k-1)})$,
gauge-covariantly averages the fermion fields over $2^4$ hypercubes on the
$(k-1)$-st lattice.
For the $k=0$ step, $D_{taste,0}=D_{taste}$ is defined in Eq.~(\ref{RGDirac}),
and $G_0^{-1} =D_{stag}+\a_0 Q^{(0)\dagger}Q^{(0)}$, where $\a_0=\a$
and $Q^{(0)}=Q$ are those introduced in Eq.~(\ref{RGDirac}).
Recall that the special $k=0$ blocking kernel is unitary; all
other blocking kernels are not.

For small momenta, $Q^{(k)\dagger}Q^{(k)}\approx {\bf 1}$,
and with $\a_k\sim 1/a_k$ it
thus follows that the eigenvalues of $G_k^{-1}$
are at least of order $1/a_k$, making $\d S_{eff}$ a short-distance
contribution to the effective gauge action.\footnote{Much smaller
  eigenvalues are allowed, as long as the corresponding eigenmodes are
  localized on a distance of at most order $a_k$.  Such modes would not
  affect the long-distance physics.
}
While this can be proved in the free case \cite{shamir05}, in the
interacting case this is an assumption that is already necessary for the conventional
RG picture to work in local, renormalizable theories.
The nature of this assumption is discussed in detail in Ref.~\cite{shamir06};
here we will assume it to be correct.  It follows that $\d S_{eff}$
remains local when we take $n_r$ to be any real number.\footnote{
  We keep $n_r$ in the range where the gauge coupling is asymptotically free.
}

The fermionic contribution to long-distance physics then resides entirely
in the $n_r$-th power of the determinant of $D_{taste,n}$ in Eq.~(\ref{RGstag}).
The problems with locality of the rooted theory originate with
taking $n_r\to 1/4$ in this power.  Our task will be to perform a faithful
replica continuation at the level of the SET.
As explained in the introduction, this is not straightforward.
Calculations in the effective theories, the SET or ChPT,
lead to explicit dependence on $n_r$ (for instance, through loops).
But there is also implicit dependence through the couplings that appear
in the effective theory, which is in general
nonperturbative, and not known.

Our strategy will be to first approximate the fourth-root theory
by a local (``re-weighted'') theory.  The fermions of this theory
do not carry a taste degree of freedom; they are taste singlets.
The multiplicity of taste-singlet fermions, $n_s$, will always be chosen
to match the fermion spectrum of the target continuum theory.
Therefore we will never have to perform any ``replica continuation'' in $n_s$;
rather, $n_s$ will always be kept a positive integer.  In our construction,
the unknown dependence of the
couplings in the effective theory on the fermions will be due
to the taste-singlet fermions only.

The fourth-root theory will be reached from the taste-singlet theory
by ``turning on'' the taste-breaking effects that introduce the nonlocal
behavior.
This is where a replica continuation away from the integers will be needed.
Because of the smallness of the taste-breaking effects,
the replica continuation will be under control.
Indeed, we will show that to any order in $a_f$, the dependence
of the taste-breaking effects on $n_r$ is polynomial,
with a degree less than the order in the $a_f$-expansion.

We start by splitting $D_{taste,n}$ into a taste-singlet part and a
taste-breaking part with vanishing trace in taste space,
\begin{eqnarray}
\label{split}
D_{taste,n}&=&D_{inv,n}+\D_n\ ,\\
D_{inv,n}&=&\tD_{inv,n}\otimes{\bf 1}\ ,\nonumber \\
\tD_{inv,n}&=&\frac{1}{4}\;\trts\left(D_{taste,n}\right)\ ,\nonumber
\end{eqnarray}
where $\trts$ denotes the trace in taste space,
and ${\bf 1}$ is the taste identity matrix.
Following Ref.~\cite{shamir06} we assume that,
in the coarse-lattice theory, $\D_n$ scales like
\begin{equation}
\label{scaling}
\|a_c\D_n\|\;\leqx\; \frac{a_f}{a_c}\ .
\end{equation}
This estimate is valid modulo logarithmic corrections
to the leading power-law scaling.
For extensive discussions of this scaling assumption, we refer to
Ref.~\cite{shamir06} (see also Refs.~\cite{bgslat06,sharpelat06}).
Here we only observe that, in any theory with integer $n_r$,
this assumption is needed to establish that unrooted staggered fermions
have the usually assumed continuum limit.
However, by exploiting the proximity of the local re-weighted theory
after a large number $n$ of blocking steps,
it was argued that the scaling (\ref{scaling})
is also valid in theories with fractional $n_r$.
In this paper, we will assume this to be the case.

Using this split, we generalize the determinant in Eq.~(\ref{RGstag}) to
\begin{subequations}
\label{generalize}
\begin{equation}
\label{generalizea}
\Det^{n_r}\left(D_{taste,n}\right)\to\Det^{n_s}\left(\tD_{inv,n}\right)
\frac{\Det^{n_r}\left(D_{inv,n}+t\D_n\right)}
     {\Det^{n_r}\left(D_{inv,n}\right)}\ ,
\end{equation}
while also replacing
\begin{equation}
\label{generalizeb}
  \bltz_n\left(n_r;\cu,\{\cv^{(k)}\}\right)
  \to
  \bltz_n\left(n_s/4;\cu,\{\cv^{(k)}\}\right)\ .
\end{equation}
\end{subequations}
The generalized theory reduces to Eq.~(\ref{RGstag})
if we set $n_s=4n_r$ and $t=1$.
This generalization has two important properties.
First, if $n_s=4n_r$ and $n_r$ assumes physically interesting
values, \ie, multiples of 1/4, then $n_s$ is an integer.
Second, when $n$ is large enough, $D_{inv,n}^{-1}\D_n$
is small enough (in an ensemble-average sense)
that we may expand
\begin{eqnarray}
\label{expand}
\frac{\Det^{n_r}\left(D_{inv,n}+t\D_n\right)}{\Det^{n_r}\left(D_{inv,n}\right)}
&=&\exp\left[n_r\Tr\log{\left(1+tD_{inv,n}^{-1}\D_n\right)}\right]\\
&=&\exp\left[-n_r\Tr\left(\sum_{\ell=1}^\infty\frac{(-1)^\ell}{\ell}t^\ell\left(D_{inv,n}^{-1}\D_n\right)^\ell\right)\right]\ .\nonumber
\end{eqnarray}

The parameter $t$ interpolates between the taste-invariant
operator $D_{inv,n}$ at $t=0$ and the (blocked) staggered operator
at $t=1$.  In addition, $t$ is a book-keeping device.
The power of $t$ is, evidently, the same as the power of $D_{inv,n}^{-1}\D_n$.
As we explain in detail in Sec.~\ref{power} below, for the construction
of the effective theories we may use the bound
\begin{equation}
\label{SETbound}
  \|D_{inv,n}^{-1}\D_n\| \leqx  \frac{a_f}{a_c}=\frac{1}{2^{n+1}}\equiv\e_n\ .
\end{equation}
(The $\sim$ sign has a meaning similar to that in Eq.~(\ref{scaling}).)
We conclude that the $t$-expansion is an expansion
in powers of $a_f$ for the taste-breaking effects.

For $t=0$, the determinant ratio (\ref{expand}) collapses to one.
The taste-invariant theory at $t=0$ is thus local for any integer $n_s$,
and independent of $n_r$.  The staggered theory is reached by expanding
as in Eq.~(\ref{expand}), eventually setting $t=1$.
The rooted staggered theory is obtained by setting
$n_r$ to a quarter-integer value.
When we construct the SET to any finite order in $a_f$, the maximal power
of $t$ will be limited by that order.\footnote{
  Note that $a_f$-dependence
  which does not involve taste-symmetry breaking may result
  from other sources besides the determinant ratio (\ref{expand}).
}
By Eq.~(\ref{expand}),  the maximal power of $n_r$ is bounded by the power of $t$.
(Because of taste-tracelessness of $\D_n$, the maximal power of $n_r$ is
in fact strictly less than the power of $t$.)
The maximal power of $n_r$ is thus (strictly) less than the order in $a_f$.
Therefore, at fixed $n_s$ and to any finite order in $a_f$,
the dependence of any correlation function on $n_r$, and thus of the SET that reproduces it, will be polynomial.
This implies that, at the level of the SET, the replica continuation in $n_r$ to quarter-integer values
will be well-defined, resulting in the ``staggered SET with the replica rule.''
What this means is the following:  We start with integer $n_r$. The effective action is then
given in terms of a set
of Symanzik coefficients which are unknown functions of $n_s$, but depend polynomially
on $n_r$ (we may already set $t=1$).
With this action, one calculates correlation functions which again depend
polynomially on $n_r$ (to any finite order in $a_f$), with $n_r$ dependence coming
from the Symanzik coefficients and from loops.  Finally, one sets $n_r=n_s/4$,
and the resulting correlation function is precisely that of the rooted staggered theory.
The following
subsections contain a more detailed argument on how this works.

We comment in passing that, for $t=1$, we may also interpolate between
the taste-singlet local theory at $n_r=0$, and the (rooted) staggered theory
at $n_r=n_s/4$ by varying $n_r$ instead of $t$.  While the two ways of moving from the taste-singlet
to the staggered theory are mathematically equivalent, we find the argument
more transparent if the transition is done by varying $t$.

\subsection{\label{hybrid} The generalized theory}
In order to define the SET we first need a complete definition of the generalized staggered
theory, coupled to sources in order to generate all correlation functions.
Returning to integer $n_r$,
the theory defined by Eq.~(\ref{generalize}) contains
$n_s$ taste-singlet fermions with Dirac operator
$\tD_{inv,n}$, $n_r$ generalized staggered fermions
with Dirac operator $D_{inv,n}+t\D_n$, and $4n_r$ ghosts with Dirac operator $\tD_{inv,n}$.  Introducing sources
$H=(\etahat,\eta,\etat)$ and $\Hbar=(\etahatbar,\etabar,\etatbar)$
for the taste-singlet, generalized staggered, and ghost fields respectively,
we define  the
partition function of the generalized theory as
\begin{eqnarray}
\label{sources}
Z_n(t,n_r,n_s;H,\Hbar)&=&\int\cd\cu\prod_{k=0}^{n}\cd\cv^{(k)}\,
\bltz_n\left(\frac{n_s}{4};\cu,\left\{\cv^{(k)}\right\}\right)\\
&&\hspace{-1.2cm}\times\ \Det^{n_s}\left(\tD_{inv,n}\right)
\frac{\Det^{n_r}\left(D_{inv,n}+t\Delta_n\right)}{\Det^{n_r}\left(D_{inv,n}\right)}\;
\exp\left[\etahatbar(\tD_{inv,n}^{-1}\times\bI_{n_s})\etahat\right]\nonumber\\
&&\hspace{-1.2cm}\times\
\exp\left[\etabar\left((D_{inv,n}+t\Delta_n)^{-1}\otimes\bI_{n_r}\right)\eta
+\etatbar(D_{inv,n}^{-1}\otimes\bI_{n_r})\etat\right]\nonumber\\
&=&\int\cd\cu\prod_{k=0}^{n}\cd\cv^{(k)}\,
\bltz_n\left(\frac{n_s}{4};\cu,\left\{\cv^{(k)}\right\}\right)\nonumber\\
&&\hspace{-1.2cm}\times\ \Det^{n_s}\left(\tD_{inv,n}\right)\;
\exp\left[-n_r\;\Tr\left(\sum_{\ell=1}^\infty\frac{(-1)^\ell}{\ell}t^\ell\left(D_{inv,n}^{-1}\D_n\right)^\ell\right)\right]\nonumber\\
&&\hspace{-1.2cm}\times\ \exp\left[\etahatbar(\tD_{inv,n}^{-1}\times\bI_{n_s})\etahat
+\etabar(\tD_{inv,n}^{-1}\otimes\bI_{4n_r})\eta
+\etatbar(\tD_{inv,n}^{-1}\otimes\bI_{4n_r})\etat\right]\nonumber\\
&&\hspace{-1.2cm}\times\ \exp\left[
\etabar\left(\sum_{\ell=1}^\infty(-1)^\ell t^\ell\left(D_{inv,n}^{-1}\D_n\right)^\ell D_{inv,n}^{-1}
\otimes{\bI}_{n_r}\right)\eta\right]\ .
\nonumber
\end{eqnarray}
Here $\bI$ stands for the identity matrix,
with dimensions as indicated by the subscript.
This is a theory with two lattice parameters, $a_c$ and $a_f$.
Alternatively, we may trade $a_f$ for the small parameter $\e_n$
of Eq.~(\ref{SETbound}). In the second expression we give the explicit expansions
in the book-keeping parameter $t$.  As explained above,
for fixed $n_s$ correlation functions
expanded to some finite power in $a_f$ are polynomial in $n_r$.
For $t=1$, $n_r=n_s/4$, and $\etahat=\etahatbar=\etat=\etatbar=0$, Eq.~(\ref{sources})
is precisely the theory of $n_s$ degenerate, fourth-rooted staggered fermions.

The generalized theory has a vector-like $U(n_s|4n_r)\times U(n_r)$
graded symmetry; $U(n_s|4n_r)$ acts on the taste-singlet and ghost fields,
and $U(n_r)$ on the
generalized staggered field.  For $t=0$ the symmetry
enlarges to $U(n_s+4n_r|4n_r)$.
The discrete symmetries include hypercubic rotations and axis reversal
\cite{gs84}.  In the staggered sector,
for $t=1$ this is augmented by shift symmetry, and (softly broken)
$U(1)_\e$ symmetry in GWL form for each flavor. The vector and axial staggered
symmetries expand to a $U(n_r)_\ell \times U(n_r)_r$ chiral symmetry group \cite{ab03}.
There is no chiral symmetry in the taste-singlet and ghost sectors,
because the GWL version of $U(1)_\e$ symmetry
mixes the taste-invariant and noninvariant parts of the blocked
staggered Dirac operator \cite{bgs06}.\footnote{We remind the reader that
$D_{inv,n}$ and $\D_n$ in Eq.~(\ref{sources}) are defined in the RG taste basis,
\cf\ Eq.~(\ref{RGDirac}), and not in the standard taste basis of Refs.~\cite{saclay,gliozzi}.}

We are now ready to discuss the SET for the generalized theory.
As long as $n_r$ is a positive integer,
the lattice theory is partially quenched but local,
and we will assume that an SET for this theory
exists in Euclidean space.\footnote{
  It is sufficient to consider the SET in Euclidean space, since we
  will postpone the continuation to Minkowski space until after
  the continuum limit has been taken \cite{bgslat06}.
}
The effective theory can be written in
terms of continuum fields $\Psi=(\qhat,q,\qt)$ and $\Psibar=(\qhatbar,
\qbar,\qtbar)$ for the taste-singlet,
generalized staggered and ghost fields, respectively, as well
as a continuum gluon field $A_\m$.  As explained above, its parameters
(the couplings multiplying each operator in the Symanzik expansion) are
polynomials in $n_r$ if we work to a finite order in $a_f$; while their
dependence on $n_s$ is unknown.
Only the $n_s$ dependence survives in the continuum limit,
where the determinant ratio~(\ref{expand}) collapses to one.\footnote{
  We observe that at nonzero $a_c$ but $a_f\to 0$, \ie,
  in the limit $n\to\infty$, the lattice action is a perfect action.
}

For general $t$, $n_s$ and $n_r$,
the fundamental cutoff is the lattice spacing
of the generalized theory, $a_c$.
The SET is the effective theory for quarks and gluons with momenta much
smaller than $1/a_c$.    However, the lattice theory contains an additional
small parameter, $\e_n=a_f/a_c$, \cf\ Eq.~(\ref{SETbound}).
It will be useful for our purposes to think of the Symanzik expansion as an
expansion in $a_f=\e_n a_c$,
with Symanzik coefficients that depend on $a_c$.\footnote{
  In the following subsection, we will argue that no negative powers of $a_f$
  can appear.
}
The effective theory can be divided into three different sectors,
corresponding to three different types of operators that can occur.
The (generalized) staggered sector consists of operators made out of
staggered fields $q$ and $\qbar$ only.
Likewise, the taste-singlet--ghost sector consists of
operators made out of the ``auxiliary'' fields $\Psihat=(\qhat,\qt)$
and $\Psihatbar=(\qhatbar,\qtbar)$ only.
Finally there is the mixed sector, where each operator is made out of
both staggered and auxiliary fields.
(Of course, all operators may contain gluon fields.)

In order to establish the validity of rSChPT in Sec.~\ref{SChPT},
we will not need to know the explicit form of the SET in full generality.
In fact, we need only consider the staggered sector of the SET.
Disregarding the auxiliary and mixed sectors,
the resulting SET, defined in terms of the quark fields $q$ and $\qbar$
and the gluon fields, is invariant under all symmetries of the generalized
staggered operator $D_{inv,n}+t\D_n$.
For $t=0$ this includes taste-replica symmetry $U(4n_r)$,
while for $t=1$ this includes the smaller group $\G_4$,
as well as softly broken $U(1)_\e$ symmetry.

For the remainder of this subsection we set $t=1$,
and thus $D_{inv,n}+\D_n=D_{taste,n}$ reduces
to the RG-blocked operator
of Eq.~(\ref{RGstag}).
Symmetries that act on the space-time coordinates often take a complicated
form under RG blocking.
In particular, shift symmetry  is realized in a complicated way.
First, the RG blocking leading to Eqs.~(\ref{RGstag}) and~(\ref{sources})
was started in the RG taste basis defined in Eq.~(\ref{RGDirac}),
and shift symmetry is thus realized as
a gauge-covariant form of Eq.~(\ref{tasteshift}).
Second, the transition to the RG taste basis was
followed by $n$ additional RG blocking steps.

The physical consequences of any exact
lattice symmetry of the underlying staggered theory, nevertheless,
cannot be lost by RG blocking.  The reason is the existence of
a pull-back mapping of every coarse-lattice operator to a fine-lattice
operator \cite{shamir06}.  For $n_r=n_s/4$, where the taste-singlet and
ghost determinants drop out, this mapping gives rise to exact equality
of corresponding observables. In other words, the coarse-lattice
observables are a subset of the original fine-lattice staggered observables.

The pull-back mapping extends to $n_r \ne n_s/4$.
Consider the expectation value of a product of
coarse-lattice staggered fermion (and gauge) fields.
By undoing the RG-blocking gaussian transformations
of the fermions,
this can be rewritten as an expectation value of a corresponding
product of fine-lattice staggered fields (that depends in addition
on the original and blocked gauge fields).
Because the Boltzmann weight of the generalized theory
contains the taste-singlet
and ghost determinants,
expectation values will not be the same as in the original staggered theory.
But since the fine-lattice symmetries are unchanged,
pulled-back coarse-lattice
observables will still transform under all the staggered symmetries.
Together with other observables constructed from the fine-lattice
staggered fields, they must fall into representations of all these symmetries.
This implies that the
physical consequences of the full set of staggered symmetries remain intact.

The $t=1$ staggered-sector SET must therefore be invariant under all the
symmetries listed in Sec.~\ref{symm}.  If we derive the SET using the taste
basis some of these symmetries will take a complicated form.  In particular,
shift symmetry will mix different orders in $a=a_f$.
But other continuum fields can always be chosen
by suitable field redefinitions such that shift symmetry resumes
the simple form of Eq.~(\ref{shiftSET}) at the level of the SET.
Moreover, a SET-level field redefinition
will also eliminate any $a_c$-dependence of the SET that originates
from the matching to the coarse-lattice interpolating fields.%
  \footnote{
  Via the pull-back, the coarse-lattice operators may be regarded as a
  particular set of interpolating fields on the fine lattice as well.
  The freedom in making field redefinitions at the level of the SET thus
  parallels the freedom, discussed in Appendix~B of Ref.~\cite{sharpelat06},
  to choose different sets of interpolating fields on the fine lattice.}
The only remaining dependence of the staggered-sector SET on $a_c$
originates at this stage from the presence of the taste-singlet and
ghost determinants in the underlying theory~(\ref{sources}).

Recall now that the group generated by the four elementary shifts $S_\m$
contains translations by $2a_f$. At the level of the SET
shift symmetry enlarges to the direct product of the group $\G_4$
and the continuous translation group. In the continuum limit $a_f\to 0$
the discrete group $\G_4$ enlarges to the full taste/replica symmetry group $SU(4n_r)$
(with $\G_4$ embedded such that it acts identically on all $n_r$ replicas).

The conclusion of the above arguments is that, for $t=1$ and for any positive
integer values of $n_s$ and $n_r$,
the generalized staggered sector of the  SET assumes exactly the same
structure,
as an expansion in the fine lattice spacing $a_f$,
as the standard staggered SET for $n_r$ staggered fields.
To order $a^2_f$, this SET is derived
in Ref.~\cite{ls99} (for $n_r=1$) and Ref.~\cite{ab03} (for arbitrary $n_r$),
and is written down explicitly in Ref.~\cite{Sharpe:2004is}.
However, the Symanzik coefficients of the staggered-sector SET
of the generalized theory are not the same functions of the parameters
of the underlying theory as in the ordinary staggered SET.
In the generalized theory, the Symanzik coefficients depend on $n_s$ and $a_c$,
parameters not present in the ordinary staggered theory.  Dependence on $n_s$
arises
because of contributions from taste-singlet loops.
In addition, the $n_r$ dependence (at fixed $n_s$)
of the Symanzik coefficients is different from that of
the ordinary staggered SET, because of contributions from ghost loops.
Indeed, the reason why the auxiliary sector was introduced in the first place,
is that---unlike the original staggered theory---the SET of the generalized
theory depends polynomially on $n_r$ to any order in $a_f$, as long as $n_s$
is held fixed.

We are now ready to make contact with the rooted theory.
In order to reach the SET of the rooted theory we hold $n_s$ fixed and
choose $t=1$.  For any $t$, we may perform the replica continuation
$n_r\to n_s/4$ in any correlation function at any given order in the loop
expansion.\footnote{
  For further discussion of the replica continuation, see Sec.~\ref{valence}.}
Indeed, because the
Symanzik coefficients are polynomials in $n_r$ to any desired order in $a_f$,
this continuation from integer values of $n_r$ is well-defined.
Now, recall that the taste-singlet
and ghost sectors of the generalized theory (\ref{sources}) cancel
(for vanishing sources) when we set $n_r=n_s/4$.
As explained above, this finally eliminates all the remaining dependence
of the staggered-sector SET on the coarse spacing $a_c$,
leaving only the dependence on the fine spacing $a_f$.
We have thus succeeded in constructing the replica-continued SET for the
original blocked theory,
Eq.~(\ref{RGstag}), for any quarter-integer value of $n_r$,
and to the desired order in $a_f$.

Putting everything together, we have
shown that the familiar staggered SET for integer $n_r$, derived to order $a_f^2$ in Ref.~\cite{ls99,ab03}, and
written down explicitly and extended to order $a_f^4$ in Ref.~\cite{Sharpe:2004is},
can be used to compute any correlation function of interest to the desired
order in $a_f$. The result
should then be replica-continued
to quarter-integer values of $n_r$. This continuation
provides the correct prescription for calculating any correlation
function in the rooted theory from the staggered SET. Of course,
in practice we will not know the precise coefficients of powers of $n_r$ in the Symanzik coefficients;
indeed in practical situations the Symanzik coefficients must be treated as unknown numbers,
to be fitted from numerical data.
However, it suffices for our argument to know that the dependence is polynomial.  When we
continue in $n_r$, we then need only continue the explicit $n_r$ dependence coming from loops,
giving a result as usual in terms of unknown Symanzik coefficients.

\subsection{\label{power} Power counting}
A cornerstone in the argument of the previous section is the expansion
in Eq.~(\ref{sources}), which is convergent if the norm of $D_{inv,n}^{-1}\D_n$
is small enough.  In this subsection, we consider this condition in more
detail.  There are two issues to be considered: the effect of insertions of
$\D_n$, as well as the size of the full object in which we expand,
$D_{inv,n}^{-1}\D_n$.

In general, the SET for a lattice
theory with lattice spacing $a$ is constructed
by matching correlation functions in an expansion in $ap$, with $p\ll 1/a$
a generic momentum, to the underlying lattice theory.  To make the matching
possible in perturbation theory, one should also take $p\gg\Lambda_{QCD}$.
The Symanzik coefficients are extracted by
computing suitable one-particle irreducible correlation functions
in the lattice theory, taking all the (nonexceptional)
external momenta to be of order $p$ \cite{symanzik}.  For the part coming
from the fermions, this amounts to expanding $D_{latt}^{-1}$
around $D_{cont}^{-1}$, namely to an expansion in
$D_{cont}^{-1}(D_{latt}-D_{cont})$, where $D_{cont}$ is the Dirac operator
for the continuum-limit theory, and $D_{latt}$ is the Dirac operator of the
lattice theory.  Because $D_{latt}-D_{cont}$ is an irrelevant operator,
we expect $\|D_{latt}-D_{cont}\|\, \ltap\, ap^2$.
Also, on dimensional grounds, $\|D_{cont}^{-1}\| \sim 1/p$.
Putting it together we conclude that
$\|D_{cont}^{-1}(D_{latt}-D_{cont})\|\sim ap$
is the relevant estimate
for the construction of the SET.  Observe that this argument
is insensitive to the long-distance physics, because the effective
infrared cutoff on the loop momenta is $p$, and by assumption
$p \gg \L_{QCD}$.  In particular, the estimates are independent
of the quark masses.

In the above argument we have implicitly assumed that the momentum
flowing through a particular (sub-)diagram is of order $p$.
This need not be true for sub-diagrams with a non-negative degree
of divergence, where all ultraviolet momenta may contribute significantly
to the loop integrals.   In general, counter terms will need to be added
in order to absorb contributions from such diagrams; in a renormalizable theory there
are only a finite number of counter terms that need to be adjusted.
Symmetries may exclude (some of) these counter terms.

Let us now study how these general considerations enter the construction
of the SET for the generalized theory (\ref{sources}).
Our starting point will be the $t=0$ taste-singlet theory.
This theory is local, because $n_s$ is integer.
In order to reach the generalized staggered theory from
the taste-singlet theory, we have to expand the propagator
$(D_{inv,n}+t\D_n)^{-1}$ around $D_{inv,n}^{-1}$, and eventually set $t=1$.
The object in which we are expanding is thus
$D_{inv,n}^{-1}\D_n$.  Since $\D_n$
is an irrelevant operator (\cf~Eq.~(\ref{scaling})), repeating
the above general arguments leads to the estimate
$\|D_{inv,n}^{-1}\D_n\| \sim a_f p$, if the momentum flowing through
the diagram is order $p$.

As noted above, we must separately consider sub-diagrams
with a non-negative degree of divergence.  The contributions
of such sub-diagrams depend crucially on the number of blocking steps $n$,
as we now explain.

Consider first what happens for $k=n=0$, namely, when we have performed
only the first special RG step that takes the fermions from the one-component
to the taste basis.  We then have $a_c=2a_f$.  When we extract
the Symanzik coefficients from a lattice calculation, the loop momenta
live on the coarse lattice.  But since the coarse and fine lattice spacings
differ only by a factor of two, the loop momentum can go as high as
$p \sim 1/a_f$.  In the divergent sub-diagrams we thus have
$\|D_{inv,n}^{-1}\D_n\| \sim 1$.  Indeed, for $a_c=2a_f$,
the generalized staggered theory will develop  $\cO(1/a_c)=\cO(1/a_f)$
mass terms, since shift symmetry and $U(1)_\e$ symmetry (for any $t\ne 1$)
are broken at the (common) lattice scale.\footnote{
  The breaking of shift symmetry is qualitatively the same
  as in the theory studied in Ref.~\cite{MW}.
}

The situation is qualitatively different
after a large number $n$ of RG steps has
been performed.  Because the lattice calculation is performed on
the coarse lattice,\footnote{
  See Ref.~\cite{shamir06} for a detailed discussion on how
  the coarse-lattice diagrammatic calculation is related
  to a calculation in the underlying fine-lattice staggered theory.
}
the maximal momentum that can flow through
any sub-diagram is now of order $1/a_c$, and one arrives at the estimate
(\ref{scaling}) for the magnitude of insertions of $\D_n$.
The estimate $\|D_{inv,n}^{-1}\D_n\| \sim a_f p$
still holds, but, what has changed is that now the maximal value
that $p$ can reach is $1/a_c \ll 1/a_f$.
The conclusion is that, for extracting
the Symanzik coefficients,
the appropriate estimate is just that of Eq.~(\ref{SETbound}):
\begin{equation}
\label{SETexpand}
\|D_{inv,n}^{-1}\D_n\|\;\leqx\;a_f/a_c \ .
\end{equation}
This estimate is valid in the taste-singlet, $t=0$ theory,
on the same grounds as for any other local theory,
and we will thus assume that it is valid nonperturbatively as well.
This is all we need, because the staggered theory is constructed
as an expansion in $D_{inv,n}^{-1}\D_n$ around the taste-singlet theory.

We end this subsections with three comments.  First,
it should be noted that, in Ref.~\cite{shamir06}, the bound
\begin{equation}
\label{bound}
\|D_{inv,n}^{-1}\D_n\|\;\leqx\;a_f/(ma_c^2)
\end{equation}
was used, with $m$ the renormalized quark mass after $n$ RG steps.
Clearly, the bound (\ref{bound}) is far weaker than (\ref{SETexpand}),
and it implies that the chiral ($m\to 0$) limit
can be taken only after the continuum ($a_f\to 0$) limit.
In Ref.~\cite{shamir06}, this was necessary in order to
place a uniform bound on the difference between any taste-singlet
correlation function and the corresponding rooted correlation function
on any (including the most infrared) scale,
thereby establishing the existence of the (correct) continuum limit
for the rooted theory.
In contrast,
assuming that the scaling~(\ref{scaling}) holds,
the bound (\ref{bound}) is much too generous for the derivation of the SET
for the generalized theory~(\ref{sources}), as we have seen above.
In particular, it follows that this SET is well-defined in the chiral limit,
as is the chiral effective
theory that can be derived from the SET. The requirement that the chiral
limit for staggered fermions
be taken after the continuum limit  \cite{sv,DURR-LIMITS,bernard04,bgss06,bgss2}
is then reproduced by calculations within staggered ChPT \cite{bernard04}.
Note that, while Ref.~\cite{bernard04} finds many standard quantities
for which the limits commute in SChPT, other quantities for
which the limits do not commute are also discussed.

Our second comment is that
the original staggered theory has no power divergences, because of shift and
$U(1)_\e$ symmetry.  This is therefore
also true for the $n$-times blocked staggered theory (\ref{RGstag}),
and for the corresponding SET.  Moreover, for large $n$, the SET
for the generalized theory (\ref{sources}) at
arbitrary values for $t\in [0,1)$ is related to the SET at $t=1$
by a convergent expansion in $t$, equivalently in $\e_n=a_f/a_c$.
The implication is that, for all $t$, the SET for the generalized theory
(\ref{sources}) has no power divergences in $1/a_f$, but only in $1/a_c$.
Examples of this are given in Sec.~\ref{examples} below.

Finally, we remark that the framework introduced here resolves a concern,
discussed in Ref.~\cite{sharpelat06}, about the
renormalizability of the rooted staggered theory.  The concern is the following:
the complete notion of renormalizability requires
not only that (infinite) counterterms can be chosen to make amplitudes finite,
but also  that the finite parts of counterterms can be chosen to bring the theory into a given
scheme.  While we know that the staggered theory is renormalizable for integer $n_r$,
for non-integer $n_r$ this notion of renormalizability requires that
the finite parts of counterterms, as well as the infinite parts, are polynomial in $n_r$
to any finite order in perturbation theory.
In Ref.~\cite{sharpelat06}, the condition on
the finite parts was introduced as an additional assumption,
{\it albeit}\/ a plausible one.
Here, such a separate assumption is unnecessary.  Under the assumptions
of the RG approach \cite{shamir06},
the taste-singlet (re-weighted) theory, defined by setting
$t=0$ in Eq.~(\ref{sources}), is a local theory of $n_s$ fermions,
that moreover becomes a perfect-action lattice theory in the limit
$a_f\to 0$, for any fixed $a_c$.
Thus one expects its renormalizability to follow straightforwardly
by standard arguments.
The rooted staggered theory
is then reached by expanding in $t$, and setting $t=1$ and $n_r=n_s/4$. Because of the
bound~(\ref{SETexpand}), the expansion in $t$ just brings in positive powers of $a_f$,
and all finite (and infinite) parts of the counterterms are unaffected for any $n_r$.
Thus the rooted staggered theory is renormalizable if the taste-singlet theory
is. In addition, the two theories have  the same counterterms.

\subsection{\label{valence} Partial quenching}
Unlike other lattice discretizations of QCD,
the continuum limit of the rooted staggered theory is, inherently,
a partially-quenched theory \cite{bg94,bernard06,sharpelat06,bgss2}.
This remains true when we consider the staggered sector
of our generalized lattice theory~(\ref{sources}) all by itself.
Let us work out the example of a target theory with  $n_s$ degenerate quarks.
Our starting point is
the generalized lattice theory with the same $n_s$,
and with $t=1$.
In order to obtain the set of all correlation functions
of the physical $n_s$-flavor theory in the continuum limit,
we need to let the combination of replica and taste indices of
the external lines assume precisely $n_s$ distinct values.
This can, for example, be accomplished by fixing the taste
index of the external legs to a single value (for example, $1$), and letting the
replica indices  take on $n_s$ values (for example $1,2,\ldots,n_s$).
Alternatively, we could use all four taste indices and only $[n_s/4]$
replica indices, where the square brackets denote rounding up to the next integer.
(In this case, unless $n_s/4$ is already an integer, not all taste indices will be used
in conjunction with each replica index.)
Many other similar choices, as well as
other types of embeddings for certain classes of physical correlation functions \cite{bernard06,bgss06},
are also possible.
Prior to the replica continuation, the lattice theory is local.
The source term in Eq.~(\ref{sources}) must accommodate
all the degrees of freedom, as specified above, that will be used in physical
correlation functions.
Therefore, we must consider only theories
where $n_r$, the (still integer!) number of staggered replicas,
is not smaller than $[n_s/4]$.

When we perform the replica continuation we set
the power of the staggered and ghost determinants
in Eq.~(\ref{sources}) to $n_r=n_s/4$.  Since we have already set $t=1$,
if we turn off all sources, the partition function of
the generalized theory reduces
to the rooted partition function,
in its RG-blocked dress~(\ref{RGstag}).
During the replica continuation of any correlation function,
by definition we hold fixed all indices of the external legs,
including in particular the replica (and taste) indices.
This means that the number of replicas in the source term of Eq.~(\ref{sources})
must stay equal to or larger than $[n_s/4]$.  The mismatch created between
the power of the staggered (or ghost) determinant and the multiplicity
of the corresponding external sources means that the staggered
sector has in itself been partially-quenched unless $n_s$ is a multiple of 4.

After the replica continuation,
the correlation functions of the EFT reproduce those of the rooted
lattice theory to the same order in $a_f$.
We stress again that the replica continuation at the level of the EFT
is well defined because, as we have shown, to any order in $a_f$
the $n_r$-dependence in the underlying lattice theory~(\ref{sources})
assumes the form of a finite-degree polynomial.

In our above example,
be it before or after the replica continuation,
the $replica \times taste$ multiplicity of the staggered fields
used to generated physical correlation functions
is equal to or larger than $4[n_s/4]$, which is to be compared
with the $n_s$ physical flavors of the target theory.  As a result,
the total number of available valence
degrees of freedom will in general exceed the physical number,
and, when we finally take the continuum limit,
the physical correlation functions will
form a proper subset of the set of all (partially-quenched)
correlation functions.\footnote{
  Correlation functions lying outside of the physical subset
  may exhibit various types of pathological behavior~\cite{bg94,bgss2}.
}
This conclusion is in fact valid for any target theory.
The only exception is a target theory in which the multiplicity of every
mass-degenerate quark species is divisible by four, in which case
the theory may be obtained in the continuum limit of an unrooted
staggered theory.

Another conclusion is that the partially-quenched representation
obtained in the continuum limit is not unique.
The only restriction is that the set of all partially-quenched
correlation functions must be large enough to accommodate
all the physical correlation functions of the target continuum theory.
With the minimal choice of replicas on the external lines, $[n_s/4]$,
the vector $replica \times taste$ symmetries are represented as a
$U(4[n_s/4]\;|\;4[n_s/4]\!-\!n_s)$ graded group on the continuum-limit correlation functions.
Had we initially allowed for $n'>[n_s/4]$ values
of the replica index on the external legs, all the physical correlation
functions of the target theory would still be reproduced
once we performed the replica continuation (followed by the continuum
limit).  But there would be more ways of embedding a given physical
correlation function in the space of all correlation functions.
Correspondingly, the $replica \times taste$ symmetries would be
represented as an $U(4n'\,|\,4n'\!-\!n_s)$ graded group.
The arbitrariness in picking a range $n' \ge [n_s/4]$
for the external-legs replica index thus entails the existence of
infinitely many partially-quenched representations in the continuum limit,
all of which share the same physical subspace.

In the rooted theory, closed (``sea-quark'') fermion loops as well as
(``valence-quark'') fermion lines
attached to external legs both originate from
the same staggered fields.  Therefore the sea and valence masses are equal,
and there is no clear-cut distinction between the sea and valence sectors.
This is a necessary condition for the emergence
of a unitary, physical subspace in the continuum limit.

In practice, it is often useful to explore unitarity-violating
correlation functions in which the valence-quark mass
is allowed to vary away from the sea-quark mass.
This situation is what
is usually referred to as partial quenching.
As we have just explained, the continuum limit of the rooted theory
is automatically a partially-quenched theory, \textit{albeit} with
equal sea and valence masses.  If it is desired to study
different sea and valence masses,
it is straightforward to add a (generalized-)staggered valence sector
to the generating functional~(\ref{sources}), by simply inserting a factor
\begin{equation}
\label{val}
\exp\left[\etabar_v\left(D^v_{inv,n}+t_v\D_n)^{-1}
\otimes\bI_{n_v}\right)\eta_v\right]
\end{equation}
into the integrand.  The superscript $v$ on $D^v_{inv,n}$
indicates that a different quark mass may have been chosen in the valence
sector.  For $t_v=1$ the valence sector has all staggered symmetries.
Again, for $a_f$ small enough, an expansion can be set up in $t_v$, just as
before.

In Eq.~(\ref{val}), $\eta_v$ and $\etabar_v$ are sources for any desired
number $n_v$ of valence (generalized) staggered fields. To avoid confusion
we stress that, even if the valence-sector source term~(\ref{val})
has been added to the generating functional~(\ref{sources}),
we cannot dispose of the original source terms.  The reason is that,
if we want to consider the SET for both sea and valence quarks,
we need sources for both in order to match the complete set of
partially-quenched correlation functions between the lattice and
the effective theory.
With the valence sector~(\ref{val}) in place, the
$replica \times taste$ symmetries  form an
$U(4n'\!+\!4n_v\,|\,4n'\!+\!4n_v\!-\!n_s)$ graded group in the continuum limit.
(Of course, these symmetries will be softly broken by unequal
sea and valence masses.)
As before, $n'$ is the 
number of distinct values of the replica index that we have allowed
for the staggered fields with sea-quark mass on the external legs.

In summary, we have seen that
partial quenching occurs at \textit{three} distinct levels.
The generalized theory~(\ref{sources}) is partially quenched to begin with,
because, to keep the taste-breaking effects under control, we had to
introduce a taste-singlet sector and a taste-invariant ghost sector.
During the replica continuation, the staggered sector undergoes a second-stage
partial quenching, created by the mismatch between the power of the determinant
and the multiplicity of the sources.  Last, if we are interested in
different valence and sea masses, we need to introduce a ``conventional''
valence sector, \cf~Eq.~(\ref{val}).

\section{\label{examples} Examples}
It is instructive to consider some aspects of the SET to second order in
$a_f$ in more detail.\footnote{
  In this section we return to the theory defined by Eq.~(\ref{sources}).
  The inclusion of valence quarks with a mass unequal to that of the sea
  quarks, as described in Sec.~\ref{valence}, is straightforward.
}
The SET can be written as an expansion in $a_f$, $t$ and $n_r$,
and thus takes the general form
\begin{equation}
\label{symeft}
S(\Psi,\Psibar,A;a_f,t,n_r)
=\sum_{i=0}^\infty\sum_{j=0}^i \sum_{k=0}^{j-1}
\;(a_f)^i\, t^j\,(n_r)^k\;S_{i,j,k}(\Psi,\Psibar,A)\ .
\end{equation}
Here we already took into account that each power of $t$ has to come with
at least one power of $a_f$, and that each power of $n_r$ has to be lower
than the power of $t$ (it cannot be equal because $\trts(\D_n)=0$).
Equation~(\ref{symeft}) is manifestly polynomial in $n_r$ to any
fixed, finite order in $a_f$.  Here we allow all types of quarks
(taste-singlet, generalized staggered and ghost) to appear on the external
legs.
The staggered sector is obtained by setting $\qhat=\qhatbar=\qt=\qtbar=0$.
The coefficients in $S_{i,j,k}$ depend on both $a_c$ and $n_s$ in all sectors.
Because of this, one cannot in general
conclude that terms linear in $a_f$ have to be multiplied
by dimension-five operators, \etc.
As already explained in Sec.~\ref{hybrid}, for $t=1$ we may assume
that a correlation function calculated in the SET
does not depend on $a_c$ if we set $n_r=n_s/4$ after the calculation.
In this section, we will consider $n_r$ integer.

Because of the way the RG-blocked theory is constructed,
for general $t$ the preferred basis for (the generalized staggered sector of)
the SET is the RG-taste basis.  Using this basis while
restricting ourselves to the (generalized) staggered sector, and to $i\le 2$,
the expansion (\ref{symeft}) takes the explicit form
\begin{eqnarray}
\label{symeftquad}
S^{quad}(q,\qbar,A;a_f,t,n_r)=&&\\
&&\hspace{-3.3cm}S_{0,0,0}(q,\qbar,A)\nonumber\\
&&\hspace{-3.3cm}+a_f\left[S_{1,0,0}(q,\qbar,A)
+tS_{1,1,0}(q,\qbar,A)\right]\nonumber\\
&&\hspace{-3.3cm}+a_f^2\left[S_{2,0,0}(q,\qbar,A)+tS_{2,1,0}(q,\qbar,A)
+t^2S_{2,2,0}(q,\qbar,A)+n_rt^2S_{2,2,1}(q,\qbar,A)\right]\ .\nonumber
\end{eqnarray}
The $n_r$-dependent term (the last term) is at this order the only one
coming from the expansion of the determinant ratio in Eq.~(\ref{sources}).
The other $t$-dependent terms come from the expansion of the staggered
source term in that equation.  We note that $S_{i,0,0}$ is taste invariant,
because of taste invariance of the $t=0$ theory.  Furthermore, $S_{2,2,1}$
is taste invariant too, because the factor of $n_r t^2$ originates
from the determinant ratio in Eq.~(\ref{sources}),
which does not affect the symmetry structure of the SET.  The taste structure
of the SET is determined by the external legs, which correspond to the source
terms in Eq.~(\ref{sources}).
Since the two allowed insertions of $\D_n$ have been ``used up'' by
the determinant ratio, only the taste-invariant part of the source term
contributes to $S_{2,2,1}$.

If we set $t=1$ then, as discussed in Sec.~\ref{hybrid},
there exist a field redefinition that brings
$S^{quad}$ to the familiar form of Ref.~\cite{ls99} for $n_r=1$, or to the form of
Refs.~\cite{ab03,Sharpe:2004is} for $n_r>1$.
In particular, the redefinition removes the terms linear in $a_f$.
The Symanzik coefficients are equal to those of Refs.~\cite{ls99,ab03,Sharpe:2004is}
if one also chooses $n_s=4n_r$, a multiple of four.
For general $n_s$ and $n_r$,
the staggered SET is that of Refs.~\cite{ab03,Sharpe:2004is},
but the coefficients are different functions of $n_r$.\footnote{
  In particular, the Symanzik coefficients of all taste-breaking
  four-fermion operators in the SET
  are independent of $n_r$ and depend only on $n_s$.
}
This form of $S^{quad}$ is the one needed for
the construction of rSChPT \cite{ab03}, which we will discuss in Sec.~\ref{SChPT}.

The taste-invariant operator $D_{inv,n}$ has no
chiral symmetry,
even when the chiral limit is taken in the underlying staggered theory,
and we would thus naively
expect a linearly divergent mass term of the form $\qbar q/a_c$.
However, for large $n$, the taste-invariant theory is close to the
theory with $t=1$ in the sense explained in Sec.~\ref{power}.
In order to deviate from the $t=1$ staggered theory,
at least one power of $a_f$, coming from an insertion of $\D_n$, is needed.  Equivalently,
the $1/a_c$ linear divergence has to be multiplied by at least one
factor of $\e_n=a_f/a_c$.
In fact, even a mass term with magnitude $\sim \e_n/a_c = a_f/a_c^2$
cannot occur.
To see this, note that we may write
\begin{equation}
\label{alt}
D_{taste,n}^{-1} = D_{inv,n}^{-1} - D_{inv,n}^{-1}\D_n D_{inv,n}^{-1} + \dots\ .
\end{equation}
This shows that the order $a_f$ difference between the $t=0$ and $t=1$ theories
has to break taste,
and therefore a taste-singlet difference has to be of order $a_f^2$.  Singlet mass
terms can thus only occur in $S_{2,0,0}$ and $S_{2,2,0}$, with opposite coefficients
such that they cancel at $t=1$.

Next, let us consider nonsinglet mass terms, \ie, terms of the form
$\qbar Kq/a_c$ with some (momentum-independent)
kernel $K$ for which $\trts(K)=0$.  At order $a_f$ a
nonsinglet mass term can only be part of $S_{1,1,0}$, because
$S_{1,0,0}$ is taste invariant.  However, staggered
symmetries at $t=1$ forbid such terms in $S_{1,1,0}$,
thus excluding this possibility.  At order $a_f^2$,
a nonsinglet mass term can only appear in $tS_{2,1,0}+t^2S_{2,2,0}$
because $S_{2,0,0}$ and $S_{2,2,1}$ are taste invariant.
Let us assume that a bilinear $\qbar Kq$ appears with
coefficient $c_1$ in $S_{2,1,0}$, and with coefficient $c_2$ in $S_{2,2,0}$.
Staggered symmetries then imply that $tc_1+t^2c_2=0$ at $t=1$,
and thus $c_1+c_2=0$.  Any nonsinglet mass term at order $a_f^2$ is therefore
proportional to $t(t-1)$.
Simply put, there has to be a factor $t$ in
order to break taste symmetry, and a factor $t-1$ to break staggered
symmetries, which include $\G_4$ and $U(1)_\e$.

In order to exclude various contributions
to the nonsinglet mass terms
in the above argument, we used the fact that mass terms cannot be introduced
or removed by field redefinitions.
As we now explain, the same is not true
for operators of dimension five or higher: they cannot be excluded
by arguments based on field redefinitions.
With the taste basis of Eq.~(\ref{symeftquad}),
we know from Eq.~(\ref{SETtaste}) that taste nonsinglet Wilson-like
dimension-five operators will already appear in $S_{1,1,0}$.
Of course, being nonsinglet, such terms
will have to vanish at $t=0$.  In addition, because of staggered symmetries,
a local field redefinition can be found removing such terms at $t=1$.  However,
this same field redefinition applied to the SET at $t\ne 1$ will, in general,
{\it introduce} taste-breaking terms at $t=0$.  So, all we can conclude is
that {\it before} the field redefinition such terms are proportional to $t$,
while {\it after} the field redefinition they are proportional to $t-1$.
We cannot conclude that they are proportional to $t(t-1)$.  In the case
of the mass terms discussed above, stronger conclusions are possible,
because dimension-three terms cannot be removed by a field redefinition.

\section{\label{SChPT} Staggered chiral perturbation theory}
In this section, we will discuss the transition from the SET to
staggered ChPT, or
SChPT.  For integer $n_r$ and $n_s=4n_r$ the derivation was first given
in Ref.~\cite{ls99} (for $n_r=1$) and Ref.~\cite{ab03} (for $n_r>1$), and we refer to those papers for details
on the explicit construction of the SChPT chiral Lagrangian.
Here we will focus on the continuation to $n_r=n_s/4$,
with $n_s$ as always a positive integer.

\subsection{\label{rSChPT} The transition to staggered chiral perturbation theory}
In the previous section we explained how the appropriate SET for a
rooted staggered theory can be constructed.  Holding $n_s$ fixed, the
Symanzik coefficients are polynomials in $n_r$, and thus
have no singularities at quarter-integer values of $n_r$.
The rooted staggered SET is obtained as a replica rule:  calculate
correlation functions to a given order in $a_f$, then set
$n_r=n_s/4$.  For the next step---the transition to ChPT---we
must again retain both $n_s$ and $n_r$ as independent variables.
In ChPT, as for the SET, the replica continuation in $n_r$
will be well-defined at fixed $n_s$, and SChPT with the replica rule,
namely rSChPT,
will be recovered after the continuation to $n_r=n_s/4$.

When we calculate correlation functions using the SET for
the generalized theory (\ref{sources}), dependence on $n_r$ occurs
in two ways: through the polynomial dependence of the Symanzik
coefficients, and through fermion loops.
Once we have calculated a certain correlation function to some
order in $a_f$ and to a given order in the loop expansion,
the dependence on $n_r$ is thus explicitly known.
Technically, this dependence will not be a polynomial, because only
the inverse quark propagators, and not the quark propagators
themselves, depend polynomially on $n_r$.
However, each quark propagator can be
re-expanded around that of the $t=0$ theory
in terms of $a_f$, and thus $n_r$, just as in
the underlying lattice theory (Eq.~(\ref{sources})).

This sets the stage for the derivation of the appropriate chiral
theory for QCD with rooted staggered fermions.
The continuum chiral theory is
an effective theory for low-energy scales where only Goldstone
bosons can appear on the external lines. It can be organized
as an expansion in $p/\L_\chi$, where $\L_\chi\sim 1$~GeV is the
chiral scale separating other hadrons from the
Goldstone bosons \cite{weinberg79}.
The chiral effective theory can be generalized to include discretization
errors, in an expansion in $a=a_f$.
The chiral effective theory is to be constructed
by matching its correlation functions to those of the underlying theory
in a double expansion in $p/\L_\chi$ and $a_f p$.
In practice,  the low-energy constants (LECs) of the chiral theory
cannot be calculated by analytic methods, and are determined by fitting
experimental or numerical data.

For positive integer $n_s$ and $n_r$, the underlying lattice theory is local,
as is the SET, and the transition to the chiral theory is more or less standard
\cite{shsi98,ls99,ab03,gss}.\footnote{
  Again, the only
element of this transition that is not absolutely standard is the assumption that
  all steps can be carried out for partially-quenched theories,
  since the generalized theory~(\ref{sources}) is partially quenched.
}
In addition, the estimate (\ref{SETexpand}) is still expected to hold,
even though it cannot be checked in perturbation theory,
because in this case the correct degrees of freedom for $p\;\ltap\; \L_\c$ are
no longer quarks and gluons.  Using the expansion (\ref{sources})
just as in Sec.~\ref{SET}, this implies
that the LECs of the chiral theory again
have to be polynomials in $n_r$.
Finally, setting $t=1$ and performing the continuation to $n_r=n_s/4$
we recover the replica-continued SChPT, or rSChPT, of Refs.~\cite{ab03,bernard06}.

We assume here that the contributions of ghosts and taste-singlet quarks in the sea
will cancel to all orders in the partially quenched ChPT once we put $n_r=n_s/4$.
All differences between the current rSChPT and the standard rSChPT \cite{ab03,bernard06}
(which does not have the taste-singlet and ghost sectors)
will then disappear in the limit  $n_r=n_s/4$, as long as we choose not to put
ghosts and taste-singlet quarks on the external lines.
Since the ghost and taste-singlet Dirac operators and masses are identical,
this cancellation is trivial at the QCD level, but not completely
trivial beyond one loop at the chiral level.%
\footnote{We thank S.\ Sharpe for emphasizing this point to us.}
We believe, though, that the cancellation
is almost certainly true order by order in SChPT, and that it will  probably be
possible to construct a ``quark flow'' proof of this.
This completes our
argument that rSChPT is the correct chiral theory for QCD with rooted
staggered fermions.

\subsection{\label{example} An example}
It is instructive to see how our approach works in a concrete example.
We will re-consider the leading-order contribution in rSChPT
to the connected scalar two-point function, previously described in detail
in Sec.~6 of Ref.~\cite{bernard06}.
Adding a scalar source $s(x)$ to the
generating functional, this two-point function
is defined as the connected part of
the second derivative with respect to this source (setting $s=0$
after taking the derivatives).
Adapting it to our generalized theory,  Eq.~(27) of Ref.~\cite{bernard06}
takes the form%
\footnote{The connection with the method and notation of  Ref.~\cite{bernard06}
is explained in Sec.~\ref{comparison}.}
\begin{equation}
\label{scalar}
Z(s)=\frac{\int\cd\cu\prod_{k=1}^n\cd\cv^{(k)}\;\bltz_n\left(\frac{n_s}{4}\right)\;
\Det^{n_r}\left(D_{taste,n}+s\otimes{\bf 1}\right)\Det^{(n_s-4n_r)}\left(\tD_{inv,n}+s\right)}
{\int\cd\cu\prod_{k=1}^n\cd\cv^{(k)}\;\bltz_n\left(\frac{n_s}{4}\right)\;
\Det^{n_r}\left(D_{taste,n}\right)\Det^{(n_s-4n_r)}\left(\tD_{inv,n}\right)}\ ,
\end{equation}
where we only indicated the $n_s$ dependence of $\bltz_n$
explicitly, \cf\ Eq.~(\ref{sources}).
Here we have chosen $t=1$, but have not yet set $n_r = n_s/4$.  It is important
to keep $n_r$ integral at this stage in order to develop the chiral theory;
keeping $n_r \not= n_s/4$ also
allows us to highlight the different ways in which
$n_s$ and $n_r$ appear.

In Eq.~(\ref{scalar}), we are starting from the fact that correlation
functions generated in the rooted staggered theory by the taste-singlet meson
source $s(x)\otimes{\bf 1}$ are
identical, in the continuum limit, to the desired correlations generated by
$s(x)$ in the target QCD theory. (See Eq.~(12) of Ref.~\cite{bgss06}.)
Note, however, that we have coupled $s(x)$ not only to the staggered quarks
but also to the ghost and taste-singlet quarks.  This keeps
the expansion in $n_r$ under control because the staggered
and ghost contributions differ only by the small taste-violating term $\D_n$.
Requiring that the taste-singlet and ghost quarks cancel
at $n_r=n_s/4$ then implies that $s(x)$ also couples to the taste-singlet quarks.

Even without a replica continuation,
the lattice theory defined by Eq.~(\ref{scalar}) is, as we discussed already above,
a partially quenched theory with $n_r$ staggered fermions, $n_s$
taste-singlet fermions, and $4n_r$ taste-singlet ghosts.  It differs
from Eq.~(\ref{sources}) in the way it is coupled to sources.  Of course,
the correlation functions that are generated by taking derivatives
with respect to $s(x)$ can also be generated by taking joint
derivatives with respect to $H(x)$ and $\Hbar(x)$ (with one each
for each space-time point).
Regardless of which type of source is used, the dynamics is that of
the sea-quark loops, and is controlled by the determinants in Eq.~(\ref{sources}).
Since in this subsection we are only
interested in the scalar two-point function, the formulation with
the source $s(x)$ is simpler.  Note that here we need the complete
effective theory, including taste-singlet and mixed sectors, because
the source $s(x)$ couples to all quarks.

At leading order in ChPT, the scalar two-point function consists
of a sum over one-loop diagrams, with pseudo-scalar mesons on the
loop (\cf\ Fig.~2 of Ref.~\cite{bernard06}).
Since $s(x)$ couples to all bilinears, staggered, taste-singlet,
and ghost, all types of pseudo-scalar mesons contribute to these
diagrams, including fermionic mesons made out of quarks and ghosts,
and mesons made only out of ghosts.
Because the taste-singlet quarks and ghost have the same Dirac operator
$\tD_{inv,n}$, the result
for the scalar two-point function that we will give below is that
of a theory with $n_s-4n_r$ taste-singlet
quarks, irrespective of the value (and in particular, sign) of
$n_s-4n_r$.  In the interest of brevity, therefore,
the discussion below will simply assume that we are
dealing with a theory with
a positive number $n_s-4n_r$ of taste-singlet quarks
(as well as $n_r$ staggered quarks).

In Ref.~\cite{bernard06} it was shown that, as expected,
in the one-flavor theory
(for which $n_s=4n_r=1$) only the non-Goldstone, heavy pseudo-scalar
taste-singlet state (the ``$\eta'$'')
contributes to this two-point function in the continuum
limit, despite the presence
of fifteen additional light pions in the underlying staggered theory.
 That this has to happen follows from the general discussion
given in Ref.~\cite{bgss06}.
Here we will not repeat the details of the calculation given in
Ref.~\cite{bernard06},
but only keep track of how the results change in the
generalized setup of the present paper, and see how $n_r$ and $n_s$ appear
in the final result.
With Ref.~\cite{bernard06}, we keep the singlet pseudo-scalar state
in the calculation for pedagogical reasons.

There are now three kinds of pions, those made out of staggered
quarks, those made out of taste-singlet quarks, and ``mixed pions,'' made out of staggered and taste-singlet quarks.
The leading-order masses of the pseudo-scalars in the staggered sector are given by
\begin{equation}
\label{stpionmasses}
M_\X^2=2\mu m+a_f^2\D_\X\ ,
\end{equation}
where $\X\in\{I,\x_\m,i\x_\m\x_\n(\m>\n),i\x_\m\x_5,\x_5\}$
labels the taste of each of the sixteen staggered
pseudo-scalars (for each replica), and the
$\D_\X$ are four LECs\footnote{$\D_{\xi_5}=0$ because this taste corresponds to the exact
Goldstone bosons.} representing the taste splittings; $m$ is the
quark mass.   Then there are pions made out of only taste-singlet quarks, with mass%
\footnote{
  The operator $\tD_{inv,n}$ has no chiral symmetry,
  and the taste-singlet quark mass is additively
  renormalized by an amount of order $a_f^2$ (see Sec.~\ref{examples}). The quantity
  $\D_{ts}$ represents the effect of this renormalization on the meson mass.
  In the case of the mixed pseudo-scalar mass,
  Eq.~(\ref{mix}), such renormalization is absorbed in $\D_{mix}$, which must be present
in any case.
}
\begin{equation}
\label{tspionmass}
M_{ts}^2=2\mu m+a_f^2\D_{ts}\ .
\end{equation}
Finally, there are mixed pseudo-scalars
made out of one taste-singlet and one staggered quark.
The mass of the latter can be parametrized, to leading order, as \cite{bbrs05}%
\begin{equation}
\label{mix}
M_{mix}^2=2\mu m+a_f^2\D_{mix}\ ,
\end{equation}
with, in general, $\D_{mix}\ne\D_{ts}$. The fact that the mass of the mixed mesons does not depend on their
staggered taste follows, as in Ref.~\cite{bbrs05}, from shift symmetry, which forbids taste-violating staggered
bilinears, and therefore forbids taste-violating four-quark operators with one staggered and
one taste-singlet bilinear.
Note that all the above masses (in particular, $M_I$) are the pseudo-scalar masses before including the
effect of the anomaly.

The LECs $\mu$,
$\D_\X$, $\D_{ts}$ and $\D_{mix}$ have unknown dependence on $n_s$, but do not depend on $n_r$.
For $\mu$ this is obvious, because it is a continuum LEC, and the
continuum theory does not depend on $n_r$ at all, but only on $n_s$.
(Recall that, in the continuum limit,
the determinants ratio~(\ref{expand})
goes to one.)  Because $\D_\X$ represents an order $a_f^2$ effect,
it can, according to our general arguments, be at most linear in $n_r$.
In practice, it is independent of $n_r$, because symmetry-breaking terms
of order $a_f^2$ in the SET do not originate from the determinant ratio
but only from the source term in Eq.~(\ref{sources}) (\cf\ the discussion
below Eq.~(\ref{symeftquad})); similar arguments apply for $\D_{mix}$
and $\D_{ts}$.
At higher order there will be $n_r$-dependent corrections
to Eqs.~(\ref{stpionmasses}) through~(\ref{mix}) coming from insertions of the
operator $S_{2,2,1}$ in Eq.~(\ref{symeftquad}).
The taste-singlet and mixed mesons
also contribute to our scalar two-point function as long as $n_r \ne n_s/4$.

Of course, the singlet pseudo-scalar (the ``$\eta'$'') will not be a Goldstone boson. It will pick up a
mass that does not vanish in the chiral and continuum limits.  In the continuum limit,
the $\eta'$ mass is given by
\begin{equation}
\label{etap}
M_{\eta'}^2=2\mu m+n_s\frac{m_0^2}{3}\ ,
\end{equation}
where $m_0^2$ is the double-hairpin parameter (\cf\ Ref.~\cite{ab03}).

Again, since the continuum limit does not
depend on $n_r$, the parameter $m_0^2$ does not depend on $n_r$.\footnote{
There are in general corrections of order $a_f^2$, as well
as momentum-dependent contributions, to this parameter, but they
do not invalidate our conclusions.  Following Ref.~\cite{bernard06}, other hairpin
contributions of order $a_f^2$ will be ignored as well.}  Away from the continuum
limit, mixing takes place in the neutral meson sector because of different scaling violations
in $M_I^2$ and $M_{ts}^2$.    This mixing leads to the appearance of pseudo-scalar mesons
with masses $M_\pm$ given by
\begin{eqnarray}
\label{diagmasses}
M_\pm^2&=&\frac{1}{2}\left(n_s\frac{m_0^2}{3}+M_I^2+M_{ts}^2
\pm\sqrt{\left(n_s\frac{m_0^2}{3}\right)^2-2(n_s-8n_r)\frac{m_0^2}{3}a_f^2\D+a_f^4\D^2}\right)\ ,\nonumber\\
a_f^2\D&\equiv&M_I^2-M_{ts}^2=a_f^2(\D_I-\D_{ts})\ .
\end{eqnarray}
In the continuum limit,  $\D=0$ and $M_I^2=M_{ts}^2\equiv 2\mu m$, so the
expression for $M_+^2$ simplifies to Eq.~(\ref{etap}).

In order to give the expression for the scalar two-point function,  we define single-particle propagators
\begin{equation}
\label{spprop}
D_A(p)=\frac{1}{p^2+M_A^2}\ ,\ \ \ \ \ A=\Xi,\ {\rm ts,\ mix}\ ,
\end{equation}
and hairpin ``double poles''
\begin{eqnarray}
\label{hairpins}
X_{I,I}(p)&=&\frac{1}{(p^2+M_-^2)(p^2+M_+^2)}\frac{p^2+M_{ts}^2}{p^2+M_I^2}\ ,\\
X_{ts,ts}(p)&=&\frac{1}{(p^2+M_-^2)(p^2+M_+^2)}\frac{p^2+M_I^2}{p^2+M_{ts}^2}\ ,\nonumber\\
X_{I,ts}(p)=X_{ts,I}(p)&=&\frac{1}{(p^2+M_-^2)(p^2+M_+^2)}\ .\nonumber
\end{eqnarray}
For $\D=0$, all hairpin double poles
become equal, and $D_I(p)=D_{ts}(p)$.

The result for the Fourier
transform $\tG(p)$ of the scalar two-point function is
\begin{eqnarray}
\label{scalar2pt}
\tG(q)&=&\m^2\int\frac{d^4p}{(2\pi)^4}\Biggl\{
2n_r^2\sum_\Xi D_\Xi(p)D_\Xi(p+q)\\
&&+16n_r(n_s-4n_r)D_{mix}(p)D_{mix}(p+q)+2(n_s-4n_r)^2D_{ts}(p)D_{ts}(p+q)\nonumber\\
&&-8n_r\frac{m_0^2}{3}\left(D_I(p)X_{I,I}(p+q)+D_I(p+q)X_{I,I}(p)\right)\nonumber\\
&&-2(n_s-4n_r)\frac{m_0^2}{3}\left(D_{ts}(p)X_{ts,ts}(p+q)+D_{ts}(p+q)X_{ts,ts}(p)\right)\nonumber\\
&&+\left(\frac{m_0^2}{3}\right)^2\Biggl[32n_r^2X_{I,I}(p)X_{I,I}(p+q)
+2(n_s-4n_r)^2 X_{ts,ts}(p)X_{ts,ts}(p+q)\nonumber\\
&&\phantom{+\left(\frac{m_0^2}{3}\right)^2\Biggl[}+16n_r(n_s-4n_r)X_{I,ts}(p)X_{I,ts}(p+q)\Biggr]\Biggr\}\ .\nonumber
\end{eqnarray}
The explicit factors $m_0^2/3$ can be eliminated from this expression by using the relation
\begin{equation}
\label{m02}
\frac{m_0^2}{3}=\frac{1}{n_s}\left(M_+^2+M_-^2-M_I^2-M_{ts}^2\right)\ .
\end{equation}
As discussed above, if we expand out the masses $M^2_\pm$ in powers of $a_f^2$,
the $n_r$ dependence of Eq.~(\ref{scalar2pt}) is polynomial. The
$n_s$ dependence is not polynomial
because the LECs $\m$, $\D_\X$, $\D_{ts}$ and $\D_{mix}$ depend on $n_s$
implicitly in an unknown way.

Let us compare the result~(\ref{scalar2pt}) to a similar calculation,
done in the taste-singlet theory obtained by replacing
$D_{taste,n}$ with $D_{inv,n}$ in Eq.~(\ref{scalar}).
To order $a_f^2$, this corresponds to setting $M_I^2=M_{mix}^2=M_{ts}^2$.
The expression for $M_+^2$ (\cf~(5.6)) again simplifies to (5.5), except that
$2\mu m$ is replaced with $M_{ts}^2$,
because $M_{ts}^2$ may still include discretization
errors.  Instead of Eq.~(\ref{scalar2pt}) we now arrive at
\begin{eqnarray}
\label{tsscalar2pt}
\tG(q)&\to&2\m^2\int\frac{d^4p}{(2\pi)^4}\Biggl\{(n_s^2-1)\frac{1}{p^2+M_{ts}^2}\ \frac{1}{(p+q)^2+M_{ts}^2}\\
&&\phantom{2\m^2\int\frac{d^4p}{(2\pi)^4}\Biggl\{}
\ \ \ \ \ \ +\frac{1}{p^2+M_{\eta',ts}^2}\ \frac{1}{(p+q)^2+M_{\eta',ts}^2}
\Biggr\}\ ,\nonumber\\
M_{\eta',ts}^2&=&M_{ts}^2+n_s\frac{m_0^2}{3}\ .\nonumber
\end{eqnarray}
As expected, this result is $n_r$-independent.
The first term on the right-hand side is recognized as the anticipated
contribution of the $n_s^2-1$ degenerate Goldstone pions of a theory
with $n_s$ (mass-degenerate) flavors.

Replacing $D_{taste,n}$ with $D_{inv,n}$ means that
the product of determinants in the denominator of Eq.~(\ref{scalar})
collapses to $\Det^{n_s}(\tD_{inv,n})$, with a similar simplification
in the numerator. Our calculation thus explicitly demonstrates
how we may consider the
rooted staggered theory as a local taste-singlet theory with small,
nonlocal corrections of order $a_f^2$, which, to any fixed order
in $a_f$, are polynomial in $n_r$.\footnote{
  By making use of the general sources in Eq.~(\ref{sources})
  this conclusion applies to any physical correlation function
  of interest.  A by-product is that the generalized theory~(\ref{sources})
  provides a alternative framework to that
  discussed in Appendix~B of Ref.~\cite{sharpelat06} for solving the ``valence rooting'' problem.
.
}
Our example also illustrates how
the nonlocality of the rooted staggered theory manifests itself in the
low-energy EFT: while Eq.~(\ref{tsscalar2pt}) satisfies unitarity,
Eq.~(\ref{scalar2pt}), at $a_f\ne 0$, does not. This is most easily seen by noting
the presence of the minus signs multiplying various terms in Eq.~(\ref{scalar2pt}),
in what should be (in a unitary theory) a positive definite correlation function.

\subsection{\label{comparison} Comparison with Reference [14]}
The present work may be compared with the
complementary argument for the validity of rSChPT given in Ref.~\cite{bernard06}.
That argument starts from ChPT for a rooted theory with four degenerate
flavors of staggered fermions, which thus describes four mass-degenerate
quark species.  The underlying lattice theory is local, trivially, because
it contains the fourth power of the fourth-rooted staggered determinant.
Staying entirely within the ChPT framework,
Ref.~\cite{bernard06} then treats the nondegenerate case by perturbing in the quark masses.
An assumption of the analyticity of the expansion around positive quark mass
is required at this point. In addition, the replica rule
(called the ``replica trick'' in Ref.~\cite{bernard06}) needs to be introduced because the
theory becomes nonlocal as one moves away from the degenerate limit.
Finally, one of the four masses can be made so large that that
quark decouples from the chiral effective theory
(at which point it can be thought of as the charm quark).
Using an assumption about the details of decoupling, one arrives
at rSChPT for three light quarks.
The decoupling
assumption leaves a small potential loophole in the
argument of Ref.~\cite{bernard06}.  While the three-flavor
chiral theory goes over, in the continuum limit, to the
standard three-flavor chiral theory of QCD,
it is not guaranteed that the LECs have the same numerical values as in QCD.
(In the initial four-flavor case, the correctness of the LECs is guaranteed,
however.)

Here, we have started instead from the fundamental lattice theory
(in RG-blocked form) and have shown
how  rSChPT may be derived from it, {\it via}\/ the SET.
The replica rule is given definite meaning in the fundamental theory,
so its appearance in the EFTs is completely natural.
In contrast, the replica rule in Ref.~\cite{bernard06} has,
by construction, meaning only at the chiral level.
It is for that reason that
a distinction was made in Ref.~\cite{bernard06}
between the power of the staggered determinant at the QCD level, which
was called $R$, and the number of replicas introduced at the chiral level, $n_r$.
Here, because the replica rule is justified at the QCD level, we need make no
such distinction. We do however need to introduce the number of flavors
of the taste-singlet quarks, $n_s$, which affects LECs in a
nonperturbative (and hence unknown) way, in order that the $n_r$ dependence
be completely controlled (indeed, polynomial).  Thus Ref.~\cite{bernard06} and the
current work represent two different generalizations of the staggered theory.
In the limit $R=n_r=n_s/4$, the two generalizations agree.  Since this
is the limit we need to take at the end of any rSChPT calculation, it is
clear that the two versions of rSChPT give the same results.\footnote{We again are assuming
that the contributions of ghosts and taste-singlet quarks
in the sea cancel to all orders in partially quenched ChPT once there
are the same number of ghosts and taste-singlet quarks, \ie, once $n_r=n_s/4$.}

Another advantage of the present approach is that it allows us to dispense
with the assumptions about decoupling and about
the analyticity of the mass expansion.
This means that the current argument closes the loophole mentioned above.
The continuum low-energy
constants are automatically those of QCD with the correct number of flavors.

On the other hand, the current argument, based as it is on Ref.~\cite{shamir06}, inherits
the assumptions of that work.  The key assumptions have already
been mentioned in the Introduction and explained in Sec.~\ref{SET}.  They are that:
\begin{itemize}
\item[$\bullet$] The effective action $\delta S_{eff}$, generated by integrating
out fermions on finer lattices, is local.

\item[$\bullet$]
The perturbative scaling laws apply, implying that the dimension-five taste-breaking
operator $\Delta_n$ goes to zero like $a_f$ (times logarithms) in the continuum limit.  This in turn is
based on the highly plausible assumption that
the theory is renormalizable to all orders
in perturbation theory for any $n_r$.

\end{itemize}

The assumption of taste-symmetry  restoration is needed in Ref.~\cite{bernard06} too,
but only for integer $n_r$, where the scaling argument is
completely standard.  The argument of Ref.~\cite{bernard06} works entirely within the chiral
theory, and the resulting rSChPT then implies the symmetry restoration
(in the chiral sector) for the rooted case.
We also note that, in the RG framework, there is an alternative route
to establish the validity of the continuum limit while relying
only on the scaling of $\Delta_n$ in the taste-singlet (re-weighted) theory
\cite{bgslat06}. Since the latter theory is local by the first assumption,
the validity of the scaling assumption needed for the RG treatment
is very plausible.
We remind the reader that there is considerable numerical evidence for the
continuum restoration of taste symmetry in the rooted case \cite{cbmilc06,milc,FM,SP,evs}.

Both the present arguments and those of Ref.~\cite{bernard06}
rely heavily on the validity of the standard partially quenched chiral theory \cite{bg94}
for describing partially quenched fundamental theories that are local.
We also need to assume here that the SET exists for partially-quenched theories,
as long as the lattice theory is local.

The calculation of the
scalar two-point function, presented in Sec.~\ref{example}, may now
be compared to the corresponding calculation in Sec.~6 of Ref.~\cite{bernard06}.
Note that Ref.~\cite{bernard06} considers only the one-flavor case as an example,
so to make the connection, we must put $n_s=1$.
The result here, Eq.~(\ref{scalar2pt}), then corresponds directly to Eq.~(41) of Ref.~\cite{bernard06}.
We can in fact make the connection at the quark flow level:
The first two lines of Eq.~(\ref{scalar2pt}) correspond to
Figs. 3(a) and (d) of Ref.~\cite{bernard06}, the next two lines correspond to Figs. 3(b) and (c), and the last two lines correspond to Fig. 3(e).
It  is straightforward to check that, if we set $n_r=n_s/4=1/4$ in
Eq.~(\ref{scalar2pt}), and $R=n_r=1/4$ in Eq.~(41) of Ref.~\cite{bernard06}, the results
are identical.

\section{\label{conclusion} Conclusion}
In this paper we presented a theoretical argument that rSChPT \cite{ab03}
is the correct chiral theory for QCD with rooted staggered fermions.
Much evidence in favor of this claim already existed, both on the
theoretical side \cite{bernard06}, as well as on the numerical side
\cite{cbmilc06,milc,FM,SP}.  Here we showed that it is possible to extend the usual
construction of the Symanzik effective theory
and chiral perturbation theory, to the rooted staggered case.
Our arguments apply equally well to any staggered quark action that has the usual staggered
symmetries, for example
standard (unimproved) staggered \cite{ks}, Asqtad \cite{ASQTAD}, HYP \cite{HYP}, Fat7bar \cite{FAT7BAR},  or
HISQ \cite{HISQ} quarks.
The version of staggered quarks used will not effect the form of the discretization effects summarized
by the effective theory, but does effect the size of these effects, which is reflected in the size of the LECs.

The effective theories are first constructed for a
taste-singlet local theory  with $n_s$ physical fermion flavors (the $t=0$ theory of Eq.~(\ref{sources})).
The rooted, nonlocal staggered theory is then reconstructed
as an expansion in the lattice spacing of the underlying staggered theory
(\ie, $a_f$), by moving smoothly from $t=0$ to $t=1$.
In this framework, the dependence on $n_r$
is polynomial to any finite order in $a_f$ and to any finite order in the
loop expansion.\footnote{For the SET, the relevant loop expansion is the
one in fermion loops; for ChPT it is the chiral loop expansion.}
The effective theories, however, are in the first instance only known
at integer values of $n_r$, where they are fairly standard.  The polynomial dependence
on $n_r$ allows us to
to make the replica continuation of any correlation function, computed order-by-order
in the effective theory for integer $n_r$, to $n_r=n_s/4$.
Once the value $n_r=n_s/4$ is reached,
the correct correlation functions of the underlying rooted lattice theory
are recovered.

The ability to extend standard techniques for the derivation of
the SET and ChPT to rooted staggered fermions does not preclude various
sicknesses in the rooted theory at nonzero $a_f$.
Indeed, in Ref.~\cite{bgs06} we argued that the rooted theory is nonlocal
at nonzero $a_f$,
due to the taste-breaking induced splittings in hadron taste multiplets.
It is essential that the replica-continued SET and SChPT
reproduce the nonlocal behavior.  This happens because loop corrections
calculated in these theories have to be continued to noninteger
number of staggered replicas as well, and the replica-continued
amplitudes cannot be reproduced from any local Lagrangian.
An explicit example of this was worked out in
Sec.~6 of Ref.~\cite{bernard06}; we revisited this example in Sec.~\ref{example}
in our generalized framework.

It is important to list the assumptions that underlie our arguments.
The most important assumption is that QCD with rooted staggered fermions
has the desired continuum limit.  This conclusion, in turn, is
based on a number of technical and testable assumptions, as explained in
detail in Ref.~\cite{shamir06} (see also Refs.~\cite{bgslat06,sharpelat06}).
If this conclusion
were to turn out to be incorrect,
that would also
invalidate the analysis presented here.  Turning this around, we consider
the success of fitting high-precision numerical results with rSChPT as direct
evidence that the conclusion of Ref.~\cite{shamir06} is, in fact, valid.

In order to keep the replica continuation under control, in Eq.~(\ref{sources})
we temporarily treated the number of dynamical quarks in the
theory ($n_s$) and the power of the staggered determinant
($n_r$) as independent.  Because $4n_r$ ghosts are needed,
we also have to assume that
the construction of the SET and ChPT goes through in the standard way
for partially-quenched (but local) theories.  This second assumption is
very common in applications of EFTs to lattice QCD.
However, one should keep in mind that, while partially quenched
ChPT \cite{bg94} is by now standard, its
foundations are not as firm as for ordinary, unquenched, ChPT.
See Ref.~\cite{Sharpe:2006pu} for a discussion of this point.

A third assumption is the technical observation that $D_{inv,n}^{-1}\D_n$
has to scale as $a_f p$, with $p$ the momentum scale at which a correlation
function in the effective theory is matched to the underlying theory.
An exception are short-distance contributions coming from sub-diagrams
with non-negative degree of divergence in which $D_{inv,n}^{-1}\D_n$
can become as large as $a_f/a_c$ at most.  The end result, the estimate
(\ref{SETexpand}), is crucial for establishing that the $n_r$-dependence
of the generalized theory (\ref{sources}) is polynomial,
to any finite order in $a_f$.%
	\footnote{We note that the same assumption, coupled with the framework introduced
	in this paper, can be used to make more plausible the argument for perturbative renormalizability of the rooted
	theory. See the discussion at the end of Sec.~\ref{power}.}
Again, we consider this assumption as noncontroversial, because it
underlies the standard derivation of EFTs for local lattice theories,
and because it is used only in the $t=0$ theory, which is local by our first
assumption.
The weaker, quark-mass dependent
bound on $D_{inv,n}^{-1}\D_n$ used in Ref.~\cite{shamir06} is not needed
for the derivation of the effective theories, and both the SET and
the chiral theory are valid in the chiral limit.  We emphasize here
that the physically sensible approach for any staggered theory (rooted or not)
is to avoid the region $m\ll a_f^2 \Lambda_{QCD}^3$, where lattice artifacts
may dominate \cite{bgss06,DURR-LIMITS,bernard04}.

In the actual construction of a SET or a chiral theory, use is made
of the symmetries of the underlying theory.  Particularly important
symmetries for staggered fermions are $U(1)_\e$ chiral symmetry and
shift symmetry, and we discussed in detail how these are realized
at the level of the SET.
Generalizing a result previously derived to order $a_f^2$ in Ref.~\cite{ls99},
we showed that for the SET, shift symmetry
enlarges to the direct product of the continuum translation group
and the finite discrete group $\G_4$.  Since this
observation holds for the SET, it also holds for any EFT derived from
the SET.
Finally, we note that our arguments also apply
to the cases of rSChPT with baryons or heavy-light mesons.

\vspace{3ex}
\noindent {\bf Acknowledgments}
\vspace{3ex}

We would like to thank Andreas Kronfeld and Steve Sharpe for helpful discussions
and for their suggestions for improving a draft version of this manuscript.
CB and MG were supported in part by the US Department of Energy.  YS was
supported by the Israel Science Foundation under grant no.~173/05.


\end{document}